\documentclass[12pt]{iopart}
\expandafter\let\csname equation*\endcsname\relax
\expandafter\let\csname endequation*\endcsname\relax
\usepackage{amsmath,amssymb,bm,dsfont}
\usepackage{cite,txfonts,comment}
\usepackage{graphicx,color}
\usepackage[colorlinks=true,linkcolor=red,urlcolor=blue,citecolor=blue]{hyperref}

\usepackage{etoolbox}
\makeatletter
\def\@mkboth#1#2{}
\newlength\appendixwidth
\preto\appendix{\addtocontents{toc}{\protect\patchl@section}}
\newcommand{\patchl@section}{%
  \settowidth{\appendixwidth}{\textbf{Appendix }}%
  \addtolength{\appendixwidth}{1.5em}%
  \patchcmd{\l@section}{1.5em}{\appendixwidth}{}{\ddt}%
}
\makeatother

\begin{document}

\title{Characterization of an operational quantum resource in a critical many-body system}

\author{S~Sarkar$^{1}$, C~Mukhopadhyay$^{2}$ and A~Bayat$^{3,4}$}

\address{Institute of Physics, Polish Academy of Sciences, Aleja Lotnikow 32/46, PL-02668 Warsaw, Poland}
\address{Quantum Information and Computation Group, Harish-Chandra Research Institute, HBNI, Allahabad 211019,  India}
\address{Institute of Fundamental and Frontier Sciences, University of Electronic Science and Technology of China, Chengdu 610051, China}
\address{Department of Physics and Astronomy, University College London, London WC1E 6BT, United Kingdom}

\ead{sarkar@ifpan.edu.pl}
\ead{chiranjibmukhopadhyay@hri.res.in}
\ead{abolfazl.bayat@uestc.edu.cn}

\begin{abstract} 
Quantum many-body systems have been extensively studied from the perspective of quantum technology, and conversely, critical phenomena in such systems have been characterized by operationally relevant resources like entanglement. In this paper, we investigate robustness of magic (RoM), the resource in magic state injection based quantum computation schemes, in the context of the transverse field anisotropic XY model. We show that the the factorizable ground state in the symmetry broken configuration is composed of an enormous number of highly magical $H$ states. We find the existence of a point very near the quantum critical point where magic contained explicitly in the correlation between two distant qubits attains a sharp maxima. Unlike bipartite entanglement, this persists over very long distances, capturing the presence of long range correlation near the phase transition. We derive scaling laws and extract corresponding exponents around criticality. Finally, we study the effect of temperature on two-qubit RoM and show that it reveals a crossover between dominance of quantum and thermal fluctuations.
\end{abstract}

\maketitle


\section{Introduction}

Many-body systems provide a rich playground for different phases of matter~\cite{sachdev2007quantum,many_body_rmp}, mediating several tasks including quantum communication~\cite{bose2003quantum,bose2007quantum}, quantum metrology~\cite{zanardi2008quantum, invernizzi2008optimal, frerot2018quantum, RevModPhys.89.035002},  remote gate implementation~\cite{yao2011robust, banchi2011nonperturbative}, and refrigeration~\cite{correa2013performance, PhysRevA.97.042124}. Ground states of many-body systems are generically highly correlated at criticality~\cite{Osterloh2002, Osborne2002}, and usually offer significant amounts of quantum resources, such as entanglement~\cite{many_body_rmp, vidal2003entanglement, bayat2012entanglement, bayat2017scaling} or quantum Fisher information~\cite{zanardi2008quantum, invernizzi2008optimal, sun2010fisher}. Most of second order quantum phase transitions are accompanied by a spontaneous symmetry breaking between two (or more) degenerate ground states at the critical point~\cite{sachdev2007quantum}, with the exception of exotic impurity quantum phase transitions~\cite{le2007entanglement, sorensen2007quantum, bayat2014order}. In practice, the breaking of symmetry is due to unavoidable environmental fluctuations. The symmetry broken ground states may show interesting properties, including the existence of points on the phase diagram of quantum many-body systems, for which the ground state is factorizable~\cite{AdessoPRL2008, AdessoPRB2009, AmicoEPL2011} into uncorrelated pure states. These points may be reached from critical points via local operations and classical communication~\cite{Vedral2012}. So far, the factorizable points are considered worth avoiding due to absence of entanglement. It would be interesting to explore whether these factorizable ground states can be useful in quantum technologies not directly relying on entanglement, which is the most celebrated resource in quantum technologies~\cite{horodecki2001entanglement, plenio2014introduction}. Many-body systems have been extensively studied for their entanglement content~\cite{many_body_rmp}, in particular, near quantum phase transitions. In fact, total quantum correlation content for the system is expected to be maximal at the critical point~\cite{Osborne2002}. However, harnessing such entanglement is practically challenging as it demands complex operations over several particles. One practically useful case is the entanglement between individual particles. Remarkably, in the transverse anisotropic XY model which is being investigated in our work, bipartite entanglement in the (thermal) ground state does not persist beyond the second nearest neighbour~\cite{Osborne2002}. One may ask whether many-body systems can provide other quantum resources which survive between individual particles, over longer distances. Whether such persistence can be used to detect the quantum critical region for a symmetry unbroken thermal state in a manner analogous to nearest-neighbour entanglement~\cite{Amico2007}, is also worth investigating.  

An important type of quantum resource is provided by magic states~\cite{bravyi} which are essential to be combined with the stabilizer gates to provide both universal and error correcting quantum computation. The concept of magic is operationally distinct from other well known quantum resources, such as, discord~\cite{Ollivier2001,Henderson2001}, coherence~\cite{baumgratz}, and symmetric quantum states~\cite{Marvian2014}. Nevertheless, the resource theory of magic~\cite{njp_magic, goursanders, campbell, howard} is intricately connected to entanglement~\cite{entmagic, entmagic2}, coherence~\cite{coherence_magical}, and especially contextuality~\cite{cont_vs_magic, contextuality_supplies_magic}, and non-Gaussianity~\cite{nong1, nong2, ferrie, marie}. Magic is a fragile resource against external noise~\cite{coherence_magical}. Hence, elaborate magic state distillation protocols have to be resorted to~\cite{bravyi, jones,du, reedmueller, factory, unified, bravyi_haah, hastings_haah, krishna_tillich}, with limitations like existence of bound magic states~\cite{campbell, magcat, howard_tight}. These protocols demand large number of unencoded magic states as raw materials to produce a few highly usable encoded magic states. 

In this paper, we investigate the phase diagram of the transverse field anisotropic XY spin chain for its single- and two-qubit magic contents. Two regimes will be considered: zero temperature in which one of the symmetry broken ground states naturally emerge as the quantum state of the system, and finite temperature regime in which degenerate symmetry unbroken ground states can equally contribute in the thermal equilibrium state.
For the ground state, we observe that the single-qubit magic lies within the stabilizer polytope throughout the disordered phase, and becomes \emph{magical} immediately after the critical point. We demonstrate the existence of two different scaling behaviors in the vicinity of this point, and extract finite size scaling coefficients. This model allows in-principle exact mining of very large number of perfect unencoded magic states at its factorized ground state and is robust against imperfect tuning of the Hamiltonian. The correlation of magic content at criticality for two qubits is shown to persist for long distances. Next, we show that the symmetry-unbroken thermal ground state also possesses magic content between distant qubits with scaling behaviour near the critical point. We then consider the magic content in the thermal states and show that thermal fluctuation eventually wipes out the magic for finite temperatures. As an application of quantification of magic in such spin systems, we furnish evidence that the quantum critical region at finite temperature can be detected by the reduced two-qubit RoM. 

The structure of the paper is as follows. In section~\ref{secII}, we review some preliminary materials. After reviewing the physical model of anisotropic XY spin chain with transverse magnetic field, we provide a brief introduction to the stabilizer computation paradigm and magic as a quantum resource. In section~\ref{secIII}, we discuss how magic content in the spin chain behaves near the critical point for the symmetry-broken ground state. In section~\ref{secIV}, we show the results for the states in thermal equilibrium. We finally conclude with a brief discussion in section~\ref{secV}.

\section{Preliminaries} 
\label{secII}

We start with a brief overview of the many-body Hamiltonian and its properties at thermal equilibrium that are analytically obtainable. This is followed by a discussion on magic as a quantum resource in the context of stabilizer computation paradigm and its quantification relevant to the many-body model at hand.

\subsection{Many-body system} 

We consider a chain of $N$ qubits with the transverse field anisotropic XY Hamiltonian~\cite{Lieb1961,Dutta2010},
\begin{align}
H = -  J \sum_{i =1}^{N-1} \left( \frac{1+\gamma}{2} \sigma_{i}^{x} \sigma_{i+1}^{x} + \frac{1-\gamma}{2} \sigma_{i}^{y} \sigma_{i+1}^{y} \right) - h \sum_{i =1}^{N} \sigma_{i}^{z} \,,
\end{align}
where $J $ is the exchange coupling, $\gamma$ is the anisotropy parameter, $h$ is the magnetic field strength, and $\sigma_i^{\alpha}$ denotes the $\alpha$-th Pauli operator, with $\alpha \in \lbrace x,y,z \rbrace $, acting on qubit $i$. As $\lambda = J/h$ varies, the ground state exhibits a second order quantum phase transition from an ordered to a disordered phase with the critical point at $\lambda = \lambda_{\text{c}} = 1$. The $U(1)$-symmetry broken ground state of this model is factorizable iff  $\lambda=\lambda_{\text{FGS}}= 1/\sqrt{1- \gamma^2}$~\cite{AmicoEPL2011}. This model can be solved analytically in the thermodynamic limit ($N \rightarrow \infty$)~\cite{Barouch1970,Barouch1971}. The quantum phase transition can be captured by the longitudinal magnetization $\langle \sigma^x \rangle$ as the order parameter of the model, which is non-zero in the ordered ferromagnetic phase for $\lambda > 1$, and vanishes in the disordered paramagnetic phase. Notably, the condition $\gamma = 1$ corresponds to the transverse Ising model, and the condition $\gamma = 0$ corresponds to the isotropic XY chain. The reduced density matrix of a qubit at site $i$ is written as $\rho_i = \frac{1}{2} \sum_{\alpha = 0,x,y,z} \langle \sigma^{\alpha}_{i} \rangle \sigma_{i}^{\alpha}$, where $\langle \sigma_{i}^{\alpha} \rangle$ is the average of $\sigma_{i}^{\alpha}$ in the ground state  ($\sigma^0 = \mathds{1}$). The two-site density matrix of qubits at sites $i$ and $j$ is written as $\rho_{ij} = \frac{1}{4} \sum_{\alpha,\beta = 0,x,y,z} \langle \sigma^{\alpha}_{i} \sigma^{\beta}_{j} \rangle \sigma_{i}^{\alpha} \otimes \sigma^{\beta}_{j}$, where $\langle \sigma_{i}^{\alpha} \sigma^{\beta}_{j} \rangle$ is the average taken in the ground state. In the thermodynamic limit, the two-point correlation functions between two qubits at distance $r$ apart for $x$- and $y$-directions are given by the determinant of the Toeplitz matrices in the following way~\cite{Pfeuty1970,Barouch1971},
\begin{minipage}{.5\linewidth}
\begin{align}
\langle \sigma_0^{x} \sigma_r^{x} \rangle  = \begin{vmatrix} G_{-1} & G_{-2} & \cdots & G_{-r} & \\ 
G_0 & G_{-1} & \cdots & G_{-r+1} & \\ 
\vdots & \vdots & \ddots & \vdots & \\ 
G_{r-2}& G_{r-3}& \cdots & G_{-1} &
\end{vmatrix} 
\,, \nonumber \end{align}
\end{minipage}%
\begin{minipage}{.5\linewidth}
\begin{align}
\langle \sigma_0^{y} \sigma_r^{y} \rangle  = \begin{vmatrix} G_{1} & G_{0} & \cdots & G_{-r+2} & \\
G_{2} & G_{1} & \cdots & G_{-r+3} & \\ 
\vdots & \vdots & \ddots & \vdots & \\ 
G_{r}& G_{r-1}& \cdots & G_{1} &
\end{vmatrix} \,,
\end{align}
\end{minipage}
where $ G_r = \frac{1}{\pi}\int_0^\pi d\phi\, \cos(\phi r)(1+\lambda\cos\phi)\frac{1}{\omega_\phi} -\frac{1}{\pi}\int_0^\pi d\phi\, \sin(\phi r) \lambda \sin(\phi)\frac{1}{\omega_\phi}$, and $\omega_\phi$ is given by $\omega_\phi = \sqrt{\gamma^2 \lambda^2 \sin^2 \phi + (1+ \lambda \cos \phi)^2}$. By assuming the two qubits to be far distant, that is, $r \rightarrow \infty$, the longitudinal magnetization $\langle \sigma^x \rangle$ can be found in a compact analytic form~\cite{Barouch1971} 
\begin{equation}
\langle \sigma^x \rangle = \sqrt{\frac{2}{1+\gamma}}\left[\gamma^2 \left(\lambda_{c}^{-2}-\lambda^{-2}\right)\right]^{\beta_x} g(\lambda)
\label{mag_x}
\,, \end{equation}
where $g(\lambda) = 1$ in the ordered phase, and vanishes in the disordered phase (see~\cite{Osborne2002} for detailed discussions), and $\beta_x = \frac{1}{8}$. As the density matrix is real, $\langle \sigma^y_i \rangle$ vanishes for all $\lambda$ and $\gamma$. The transverse magnetization $\langle \sigma_z \rangle$ is given in terms of elliptical integrals~\cite{Barouch1971},
\begin{equation}
\langle \sigma^z \rangle = \frac{1}{\pi} \int_{0}^{\pi} \frac{1+ \lambda \cos{\phi}}{\sqrt{(\gamma \lambda \sin{\phi})^2 + (1+ \lambda \cos{\phi})^2}} d\phi
\label{mag_z}
\,.\end{equation}
The two-point correlation function in $z$-direction is given by
\begin{equation}
\langle \sigma_0^{z} \sigma_r^{z} \rangle = \langle \sigma^z \rangle^2 - G_r G_{-r}
\label{mag_z}
\,.\end{equation}
These magnetizations and two-point correlators enable one to write down the reduced density matrices of the spin chain which are then used to calculate various objects of interest such as the magic content in the current work. We now go on to provide an introduction to magic as a quantum resource. 
 
\subsection{Magic state formalism}

A universal quantum computer is capable of implementing every unitary transformation on the Hilbert space of $N$ qubits. An important subset of unitary operators is the set $\lbrace U_{\text{stab}} \rbrace$ that stabilize the Pauli operators under their action, such that $U_{\text{stab}} \sigma_i ^{\alpha}U_{\text{stab}}^{\dagger} = \sigma_j^{\beta}$. These stabilizer unitaries are the key ingredients for quantum error correction~\cite{lidar_book}. Although the stabilizer unitaries can generate entanglement~\cite{cluster}, they are not universal since their action on a particular input bit-string can only cover a subset of the entire Hilbert space. In fact, a stabilizer circuit is demonstrably efficiently simulable with a classical probabilistic computer with the required depth of the classical circuit scaling polynomially with the desired precision~\cite{gk}, in contrast to an exponential scaling for a generic unitary operation. Remarkably, the injection of some extra quantum resources, known as \emph{magic} states, allows one to perform universal quantum computation, even with stabiliser circuits~\cite{bravyi}. Note that, magic is distinct from other quantum mechanical resources, e.g., entanglement. For instance, product states, or even single qudit states, can be magic states,  while certain highly entangled states, e.g., cluster states, contain no magic~\cite{cluster}. In this context, the \emph{free}, i.e., non-magical states, are called $stabilizer$ states, which are states generated from the action of  stabilizer unitaries on $|0,0,\cdots,0\rangle$, and any convex mixture. For single qubits, the stabilizer unitaries are the Pauli operators together with the Hadamard and the phase gates~\cite{njp_magic}, whose actions on $|0\rangle$ generate six pure stabilizer states. Therefore, the stabilizer states form an octahedron inscribed within the Bloch sphere~\cite{njp_magic, heinrich}, and states outside the octahedron represent magic states. For multiqubit systems, entangling gates such as CNOTs may also belong to the set of stabilizer unitaries. 

\subsection{Quantification of magic}

Quantification of any quantum resource is usually possible in terms of several different monotones. For magic, the most obvious candidates are the distance based measures, e.g., relative entropy of magic, or the trace distance of magic, which have been demonstrated to satisfy monotonicity under stabilizer operations~\cite{njp_magic}. However, computations of these quantities involve optimizations over the entire Hilbert space, and hence become prohibitively hard for all but very small systems. We mention in this connection that in case of an arbitrary number of odd dimensional qudits, magic monotones based on the negativity of the discrete Wigner function do not require any optimization~\cite{njp_magic}. However, in our case, where constituent spins are two-dimensional, such monotones do not exist. 

Thus, to efficiently quantify magic, we use the \emph{Robustness of Magic} (RoM)~\cite{howard, heinrich, goursanders}, defined for an arbitrary quantum state $\rho$ as,
\begin{align}
R(\rho) := \inf_{S_k \in S} \left\lbrace m \geq 0 \ :  \frac{\rho + mS_k}{1+m} \in S  \right \rbrace \,,
\end{align}
where $S$ is the set of stabilizer states. The reader may note that the formulation of the RoM is in terms of a semi-definite program, which makes it far more efficient to compute than distance based magic monotones. The RoM $R(\rho)$ quantifies the minimal weight of stabilizer states which, upon mixture with $\rho$, yields a stabilizer state. Using the  overcompleteness of stabilizer states, one may express $\rho$ as a pseudomixture of stabilizer states $ S_k \in S$ such that $\rho = \sum_{k} X_k S_k$, where the weights $\lbrace X_k \rbrace$ are in general, arbitrary real numbers satisfying the normalization constraint $\sum_{k} X_k = 1$. For a stabilizer state $\rho$, it is possible to find at least one such decomposition where all $\lbrace X_k \rbrace$s are non-negative. However, if $\rho$ is not a stabilizer state, at least one of the  weights $X_k$ is negative. The RoM can be alternatively expressed as the following linearized minimization problem~\cite{campbell},
\begin{align}
R(\rho)= \inf_{\{X_k\}} &\left\{ \sum_k |X_k| - 1 :   AX=B \right\} 
\,, \label{Magic_def} 
\end{align}
where $A_{\alpha \beta}=\text{Tr}(\sigma^\alpha S_\beta)$ and $B_{\alpha}=\text{Tr}(\sigma^\alpha \rho)$, with $\sigma^\alpha$ as the $\alpha$-th Pauli operator. This linear programming problem can be solved numerically via any convex optimization package. It is worth emphasizing that the maximum RoM for a single qubit confined to one of the equatorial planes of the Bloch sphere is $\sqrt{2}-1$, attained by the $H$ states~\cite{howard,heinrich}. For the spin chain we are working with, the single-qubit RoM can be written in the following simple form (see \ref{Appendix A})
\begin{equation}
R_{\gamma} (\lambda) = \text{max}\left[\langle \sigma^x \rangle + \langle \sigma^z \rangle -1 , 0 \right].
\label{robustness_simple_Expr}
\end{equation} 
In the extreme regime of the disordered phase,  i.e., $\lambda \rightarrow 0$, all the spins point towards the magnetic field, and show no magic. Deep in the ordered phase $\lambda \rightarrow \infty$,  the ground state takes a GHZ form $\frac{1}{\sqrt{2}}\left( |0,0, \cdots ,0\rangle - |1,1, \cdots, 1 \rangle \right) $~\cite{Osborne2002}, and thus every single qubit is maximally mixed, and hence non-magical. Near the critical point however, the ground state is highly entangled, and thus one may wonder whether any magic would emerge in the quantum state of the system.

\section{Magic in symmetry-broken ground state}
\label{secIII}

We first engage in computing the RoM in the ground state of the spin chain where the the global phase flip symmetry (invariance of the Hamiltonian under the action of the operator $U=\prod_i \sigma^z_i$) is spontaneously broken. In a physical realization this usually occurs due to small external perturbations. Therefore, in the ordered phase, where there is a two-fold degeneracy, we choose the ground state with positive order parameter, without loss of generality in the results obtained. The analyses of the behaviour of RoM near criticality for single-qubit and two-qubit subsystems are presented in the following.  

\subsection{Single-qubit magic analysis} 

The reduced density matrix for a single qubit in the symmetry broken ground state is $\frac{1}{2} ( \mathds{1} + \langle \sigma^{x} \rangle \sigma^{x} + \langle \sigma^{z} \rangle \sigma^{z} )$. When $\lambda$ is close to the critical point $\lambda_c$ in the ordered phase, the order parameter $\langle \sigma^x \rangle$, given by Eq.~\eqref{mag_x}, scales approximately as $\langle\sigma^x\rangle = K_x (\lambda - \lambda_c)^{\beta_x}$, where $K_x$ is a constant, expressible in terms of the anisotropy parameter $\gamma$~\cite{Osborne2002}. The derivative of the transverse magnetization has a logarithmic divergence near criticality, but $\langle\sigma^{z}\rangle$ also can be numerically approximated by the algebraic behaviour, $\langle \sigma^z \rangle \simeq  \langle \sigma^z \rangle_{c} + K_z (\lambda - \lambda_c)^{\beta_z} $, where $\langle \sigma^z \rangle_{c}$ is the transverse magnetization at the critical point, and $\beta_z$, $K_z$ are $\gamma$-dependent constants. It is noteworthy that $\beta_z \gg \beta_x$ for all possible values of $\gamma \in (0,1]$ (see \ref{Appendix B}).

For every anisotropy parameter $\gamma$,  one can locate a value of $\lambda = \lambda_c^{*} (\gamma)$, such that the RoM vanishes for every $\lambda \leq \lambda_c^{*}$, and finite thereafter. We call this point $\lambda_c^{*}(\gamma)$ as the \emph{magic pseudocritical point} (MPP).  Although $\lambda_c^{*}(\gamma)$ is very close to the critical point, it always lies in the ordered phase. Note that the existence of an MPP is a consequence of the definition of RoM in Eq.~\eqref{robustness_simple_Expr}, and does not imply a new criticality. Nonetheless, we call  $\lambda_c^{*}$ the MPP, since $R_{\gamma}(\lambda)$ exhibits power law scaling behaviour in the vicinity of $\lambda_c^{*}$, as shown below. By inserting the algebraic behaviours of $\langle\sigma_x \rangle$ and $\langle \sigma_z \rangle$ in Eq.~\eqref{robustness_simple_Expr}, and by using the constraint $R_{\gamma}(\lambda_{c}^{*}) = 0$, we get
\begin{equation}
 R_{\gamma} (\lambda) =   K_x \left( \left(\lambda - \lambda_c\right)^{\beta_x} - \delta\lambda_c^{\beta_x} \right) + K_z \left( \left(\lambda - \lambda_c\right)^{\beta_z} - \delta\lambda_c^{\beta_z} \right)  ,
 \label{RoM_scaling}
\end{equation}
where $\delta \lambda_c = \lambda_c^{*} - \lambda_c$  is the distance between the MPP and the critical point. The above result is valid for $\lambda > \lambda_{c}^{*}$. When the parameter $\lambda$ is in very close vicinity of the MPP,  determined by $\lambda - \lambda_{c}^{*} \ll \delta \lambda_{c}$, keeping terms up to first order in $(\lambda - \lambda_{c}^{*}) / \delta\lambda_c$ in the expression of RoM in Eq.~\eqref{RoM_scaling} results in the following linear scaling behaviour about the MPP
\begin{equation}
R_{\gamma} (\lambda) =  T_{\gamma} \left(\lambda - \lambda_c ^{*}\right) , \hspace{0.3in} \text{for } \lambda - \lambda_{c}^{*} \ll \delta\lambda_c, 
\label{linear_scaling}
\end{equation}
where the prefactor is given by $T_{\gamma} = K_x \beta_x (\delta\lambda_c)^{\beta_x -1} + K_z \beta_z (\delta\lambda_c)^{\beta_z -1} $. Beyond this regime, when $\lambda - \lambda_{c}^{*} > \delta\lambda_c$, the RoM has contributions from two algebraic behaviours, as given in Eq.~\eqref{RoM_scaling}, up to a constant which depends on $\gamma$. As $\beta_x$ is smaller than $\beta_z$ the dominant behaviour of the RoM stems from the first term in Eq.~\eqref{RoM_scaling}, i.e., $R_{\gamma} (\lambda) \simeq K_x \left(\lambda-\lambda_c\right)^{\beta_x} + \text{const} $. Hence, the derivative of RoM scales as  
\begin{equation}
 \frac{\partial R_{\gamma} (\lambda)}{\partial \lambda} \propto \left(\lambda - \lambda_c \right)^{\beta_x -1} \hspace{0.1in}. 
\label{derv}
\end{equation}
We note that this expansion is about the critical point, and as such, valid for all $\lambda > \lambda_{c}^{*}$ reasonably near criticality. 

\subsubsection{Transverse Ising chain}

\begin{figure}[t]
\centering
  \begin{tabular}{ccc}
    \begin{minipage}{0.3\linewidth} 
    \includegraphics[width=0.98\linewidth]{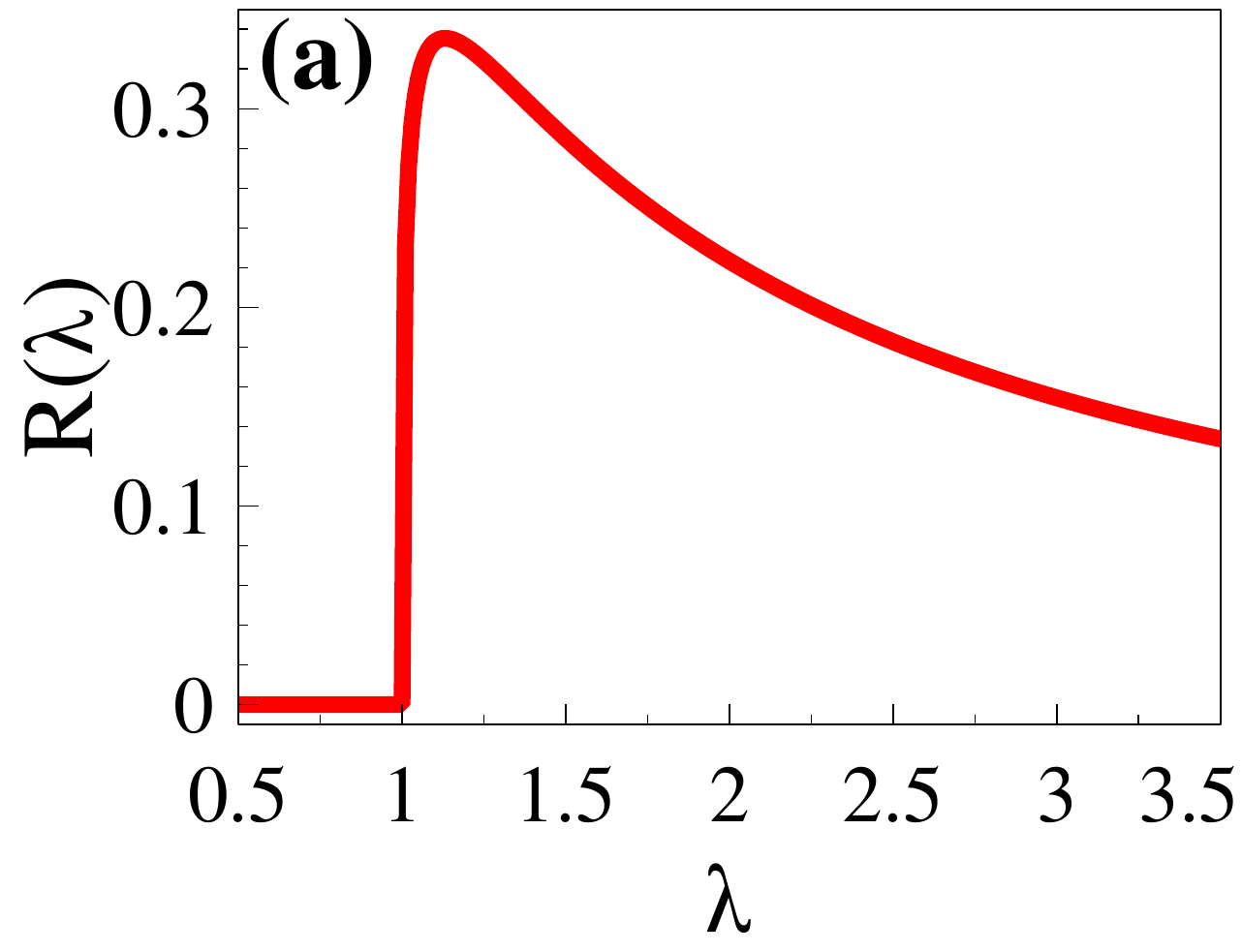} 
    \end{minipage} &
     \begin{minipage}{0.3\linewidth} 
    \includegraphics[width=0.98\linewidth]{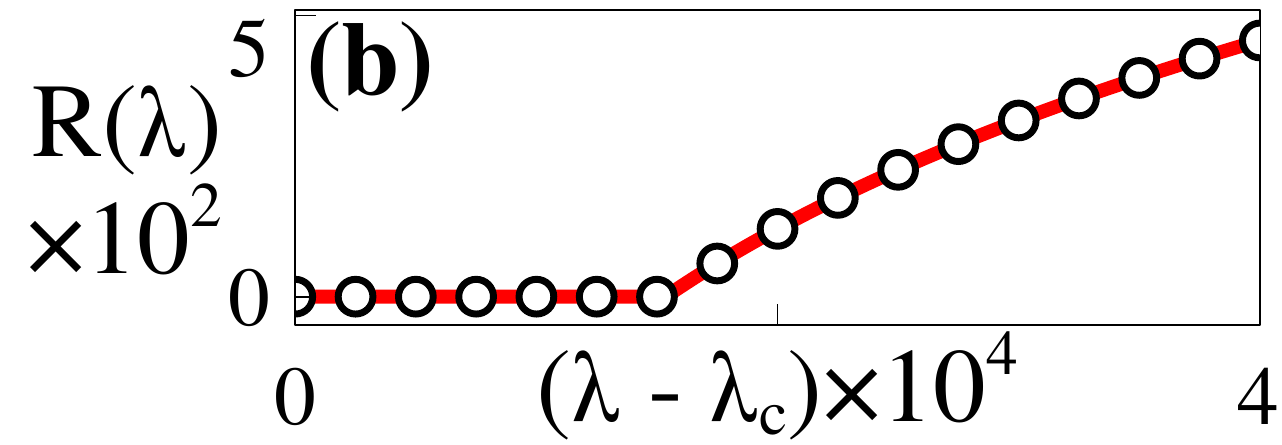} \\ 
    \includegraphics[width=0.98\linewidth]{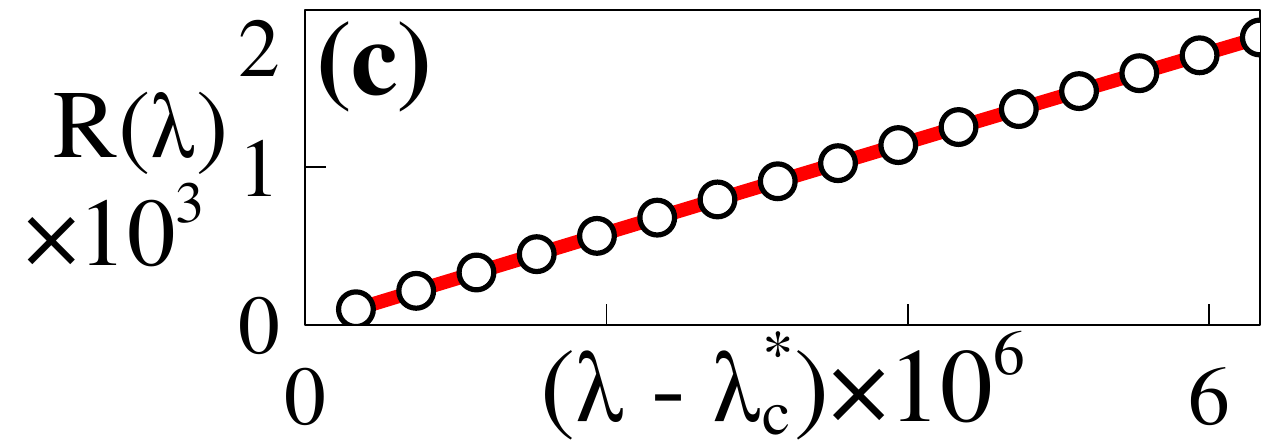} 
    \end{minipage} &    
    \begin{minipage}{0.3\linewidth} 
    \includegraphics[width=0.98\linewidth]{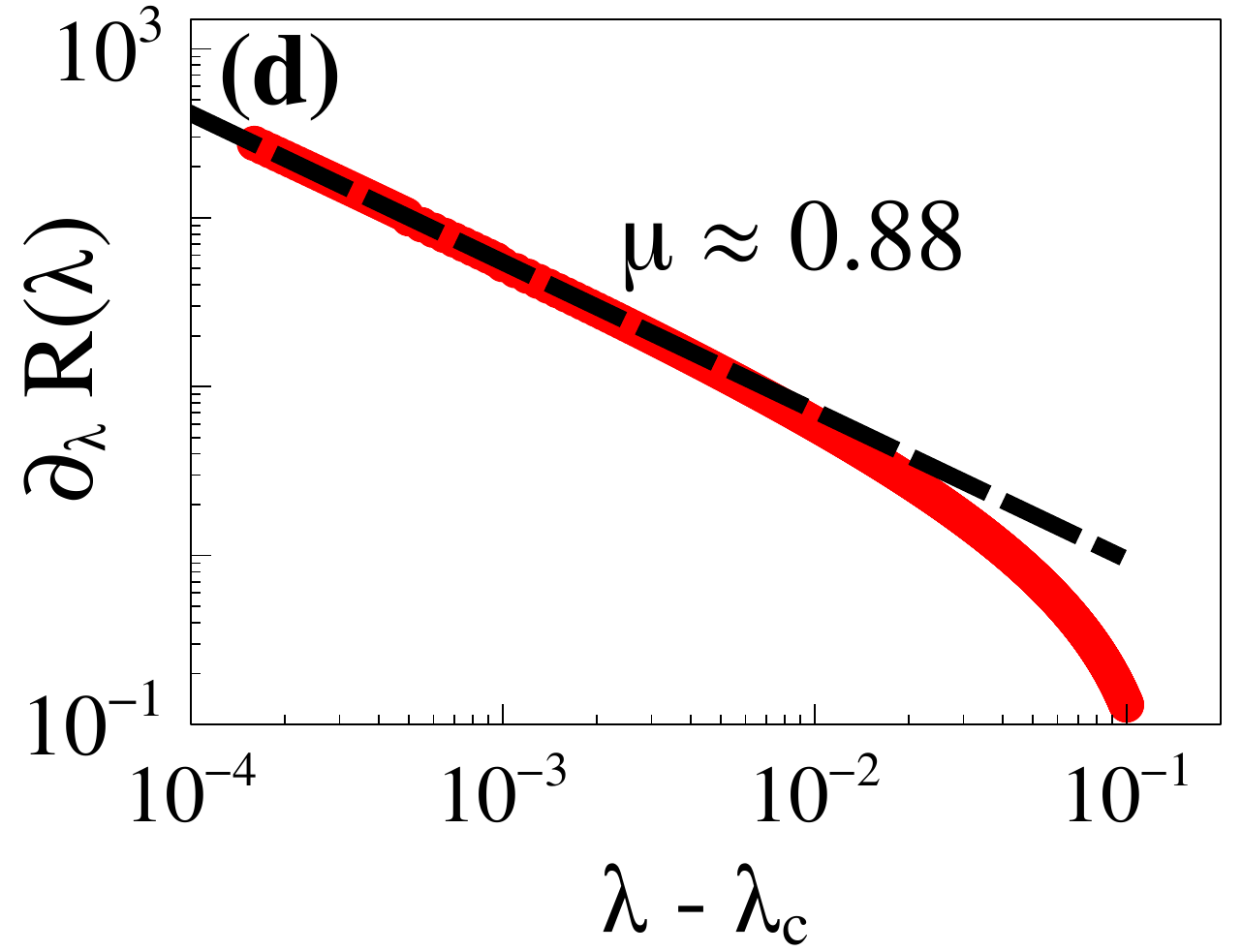} 
    \end{minipage}  
  \end{tabular}
\caption{\textbf{Infinite transverse Ising chain}. (a) RoM for the ground state of infinite transverse Ising chain plotted with $\lambda$. (b) RoM rises at the MPP which is slightly after $\lambda_{c}$. (c) Linear scaling of RoM near the MPP. (d) The derivative of RoM plotted against deviation from criticality in a log-log plot. }
\label{TI_plot}
\end{figure}
  
We first focus on the transverse Ising chain ($\gamma=1$) in the thermodynamic limit. Using the analytic form of $\langle \sigma^x \rangle$ from Eq.~(\ref{mag_x}) and numerically computing $\langle \sigma^z \rangle$ from  Eq.~(\ref{mag_z}) one can compute $R_{\gamma{=}1}(\lambda)$ using Eq.~\eqref{robustness_simple_Expr}. In figure~\ref{TI_plot}(a), we plot the RoM $R_{\gamma{=}1}(\lambda)$ as a function of the control parameter $\lambda$. As figure~\ref{TI_plot}(b) shows, the rising of RoM from zero, namely MPP,  is very close to the critical point, and takes place at around $\lambda_{c}^{*} =1.00015$. After the MPP, the RoM rises very steeply and reaches its maximum around $\lambda = \lambda_{\max}=1.13$, and then decays slowly as $\lambda$ increases further. Since the MPP is very close to the critical point, the region for linear scaling of magic, as given by Eq.~\eqref{linear_scaling}, is very small, and displayed in figure~\ref{TI_plot}(c). In figure~\ref{TI_plot}(d), we plot the derivative $\partial_{\lambda} R_{\gamma{=}1}(\lambda)$ as a function of $\lambda - \lambda_c$ in the log-log scale, which shows algebraic behaviour of the form $\partial_{\lambda} R_{\gamma{=}1}(\lambda)\sim (\lambda - \lambda_c)^{-\mu} $, with $\mu=0.88$. The fitting parameter $\mu$  is very close to $1-\beta_x$ (with $\beta_x=1/8$), giving an excellent agreement with the prediction of Eq.~\eqref{derv}.  

\subsubsection{Anisotropic XY chain} 

\begin{figure}[t]
\centering
  \begin{tabular}{cc}
    \begin{minipage}{0.35\linewidth} 
    \includegraphics[width=0.9\linewidth]{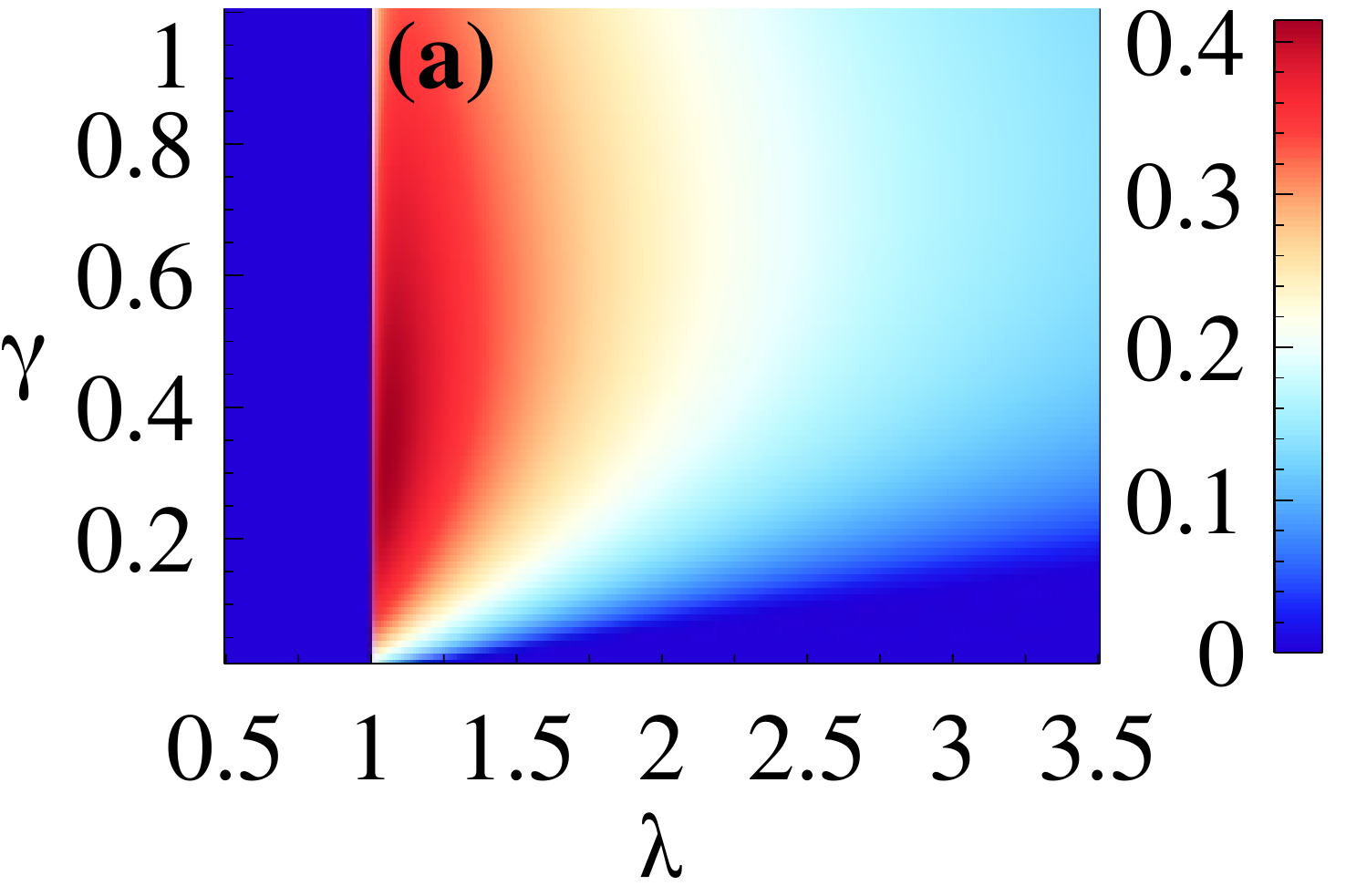} 
    \end{minipage} &
     \begin{minipage}{0.35\linewidth} 
    \includegraphics[width=0.9\linewidth]{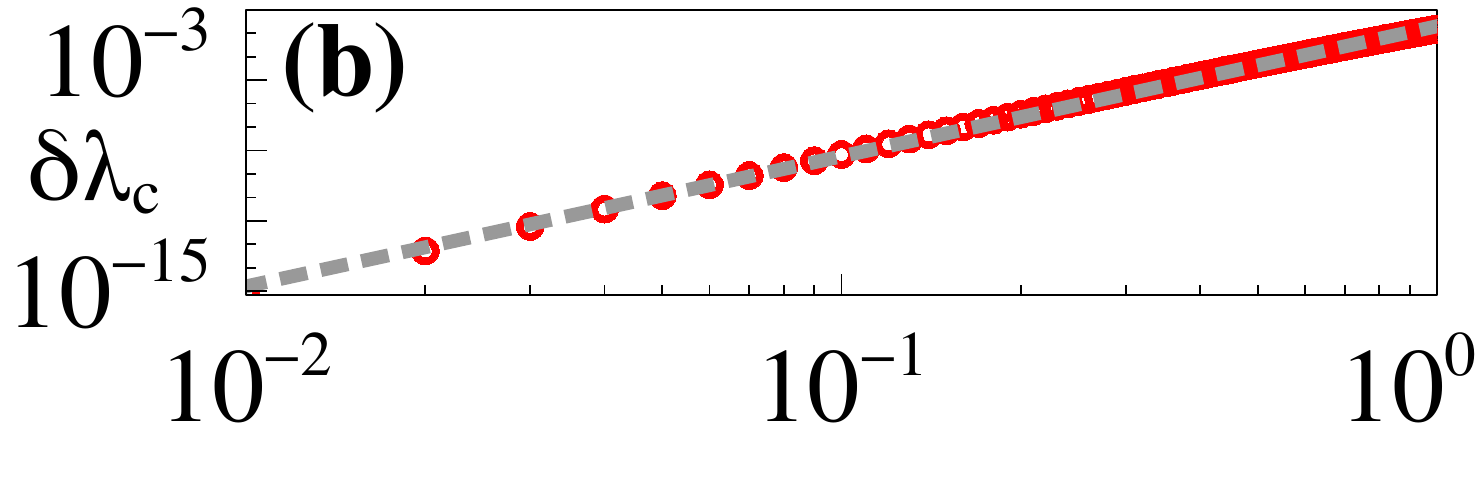} \\ 
    \includegraphics[width=0.9\linewidth]{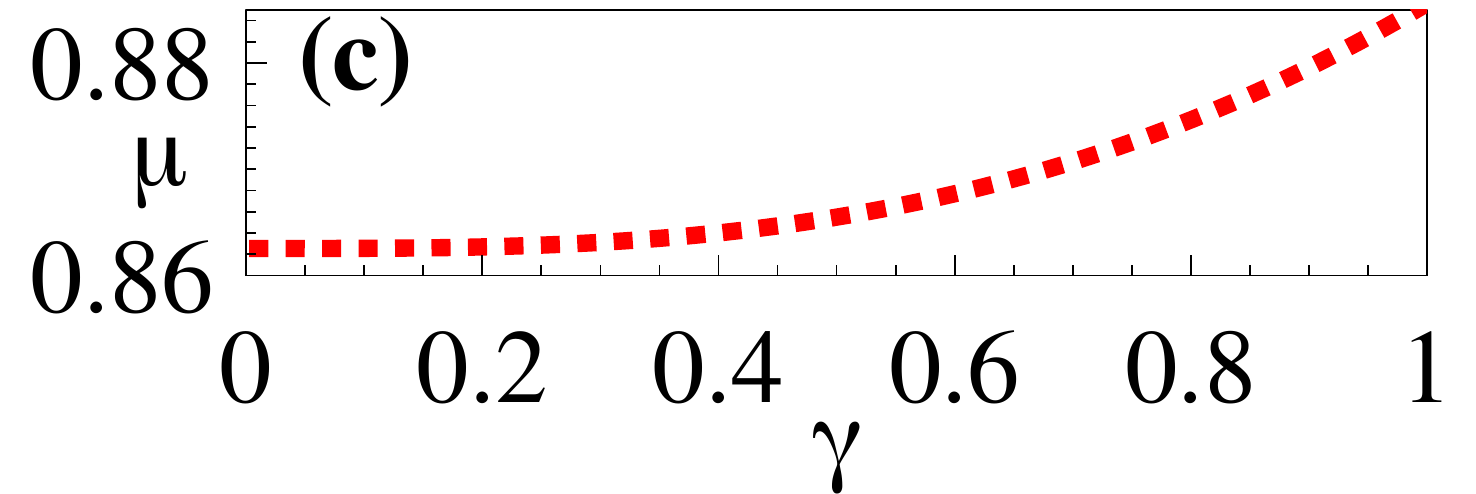} 
    \end{minipage} \\    
    \begin{minipage}{0.35\linewidth} 
    \includegraphics[width=0.9\linewidth]{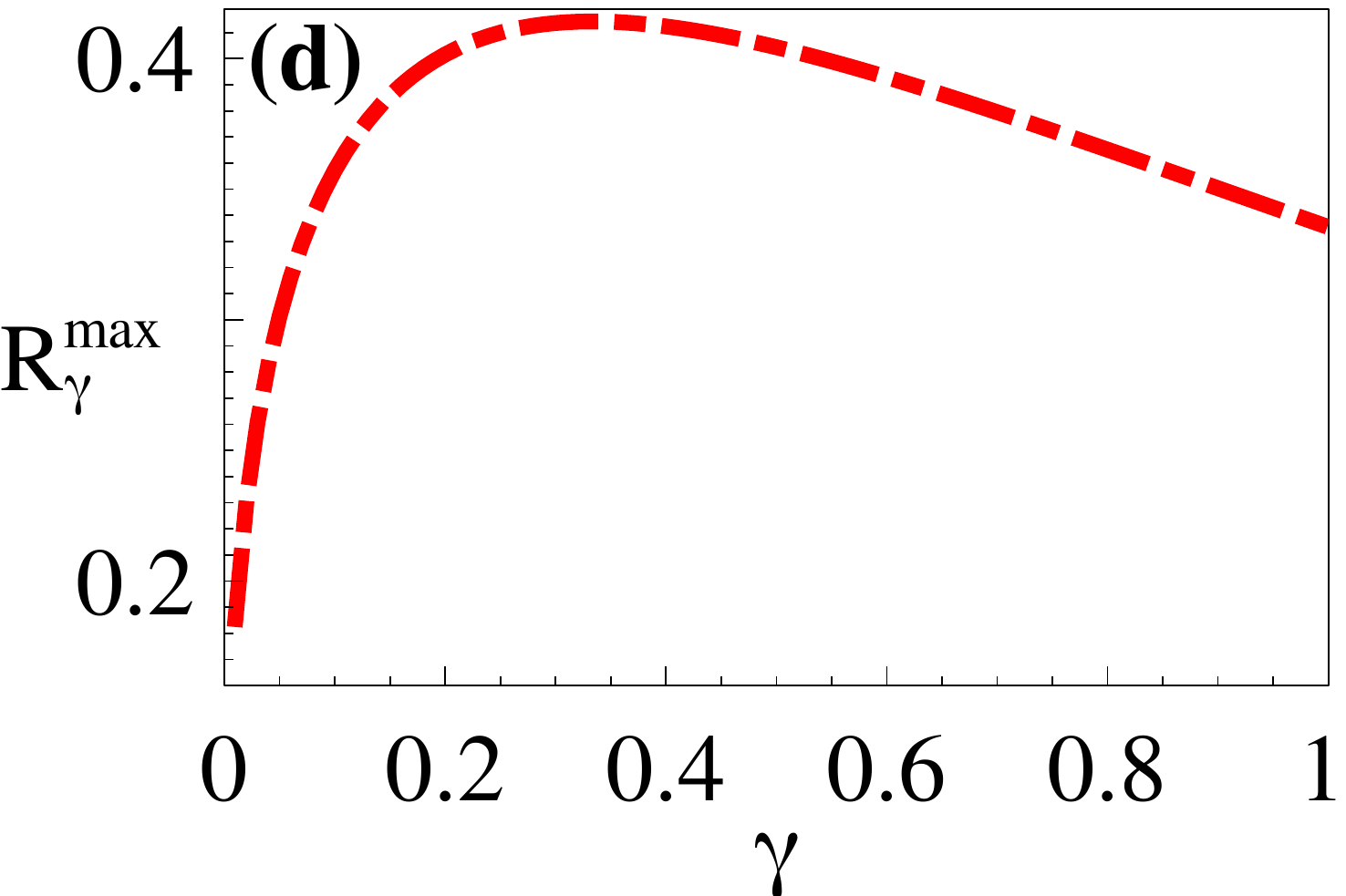} 
    \end{minipage}     & 
    \begin{minipage}{0.35\linewidth} 
     \includegraphics[width=0.9\linewidth]{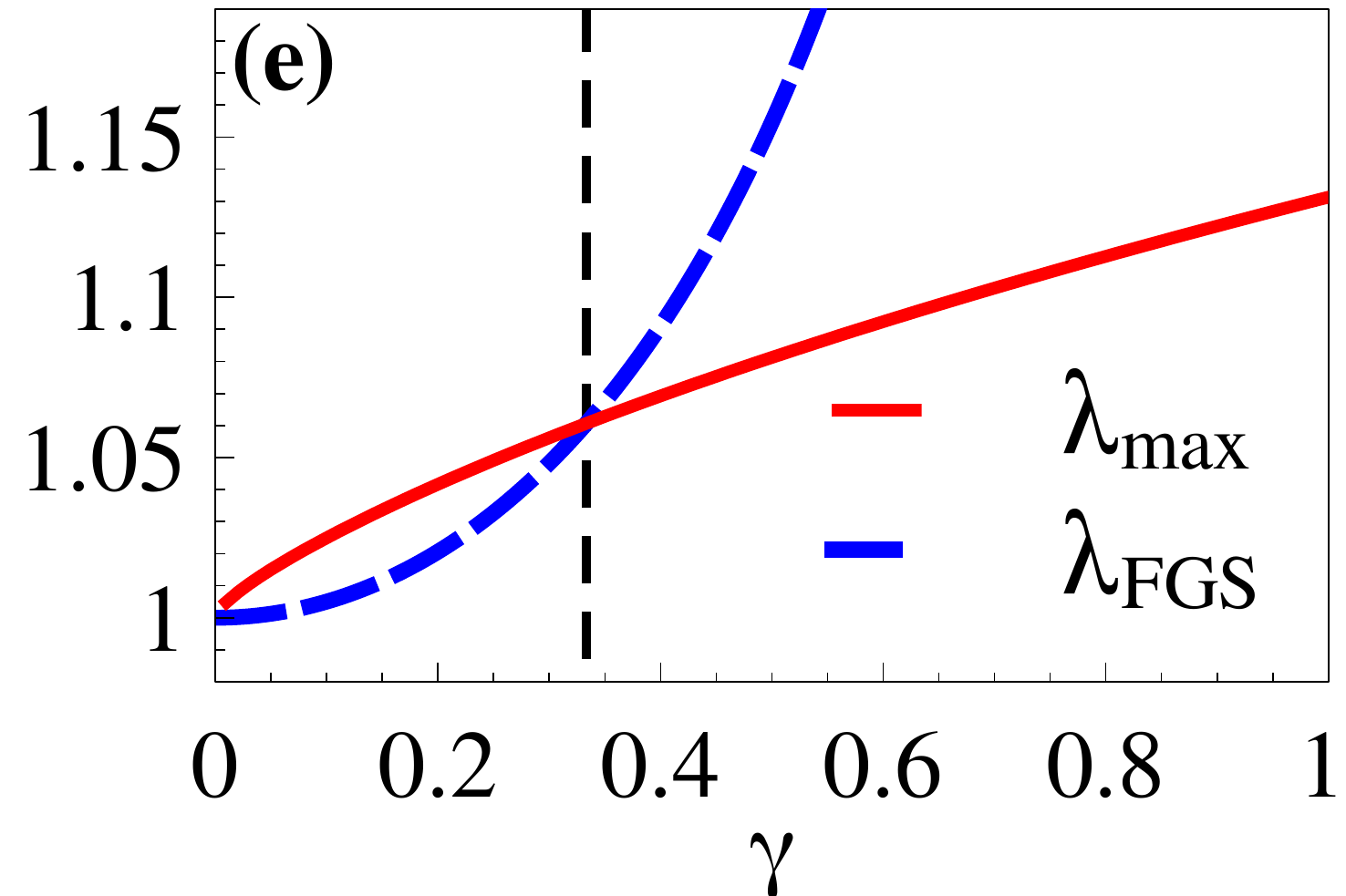} 
    \end{minipage}
\end{tabular}  
\caption{\textbf{Infinite transverse XY chain}. (a) Density plot of RoM at ground state as a function of $\lambda$ and $\gamma$. (b) Deviation of the MPP $\lambda_c^{*}$ from actual critical point $\lambda_c$ for different anisotropy parameters (red circles), which shows a scaling behaviour with exponent $\approx 5.55$ (gray dashed fitting line).  (c) Critical exponent $\mu$ calculated for different $\gamma$'s. (d) Maximum RoM  attainable vs. $\gamma$. (e) The values $\lambda_{\text{max}}$ (red solid line) and $\lambda_{\text{FGS}}$ (blue dashed line) intersect exactly where the global maxima of RoM is reached (black broken line). }
\label{XY_plot}
\end{figure}

We now consider the more general case of the anisotropic $XY$ chain, where $\gamma \in (0,1]$. In figure~\ref{XY_plot}(a), we plot $R_{\gamma} (\lambda)$ as a function of both anisotropy parameter $\gamma$ and the control parameter $\lambda$. Again, for a fixed value of $\gamma$, as $\lambda$ varies, there is a sharp rise in the RoM right after the MPP until it reaches its maximum $R_{\gamma}^{\max}=R_{\gamma}(\lambda=\lambda_{\max})$, and then decays gradually. In  figure~\ref{XY_plot}(b), we plot the MPP as a function of $\gamma$. As the figure shows, $\lambda_c^{*}$ remains very close to $\lambda_{c}$, and monotonously moves towards $\lambda_c = 1$, as the anisotropy in the Hamiltonian decreases. Interestingly, the deviation $\delta\lambda_c$ shows power law scaling with $\gamma$, which is, $\delta\lambda_c \sim \gamma^{5.55}$. 

Approaching the isotropic limit as $\gamma \rightarrow 0$,  magic in the system vanishes for any value of the control parameter $\lambda$, as the longitudinal magnetization $\langle \sigma^{x} \rangle$ vanishes throughout. The corresponding scaling exponents $\mu$ is plotted for various anisotropy parameters $\gamma$ in figure~\ref{XY_plot}(c) which shows small variations near $1-\beta_x$. The slight variation of $\mu$ across the phase diagram is due to sub-dominant corrections from $\langle \sigma^z \rangle$. In figure~\ref{XY_plot}(d), we plot the maximum RoM $R_{\gamma}^{\max}$ as a function of anisotropy parameter $\gamma$. Interestingly, the $R_{\gamma}^{\max}$ peaks at $\gamma = \gamma_0= 1/3$, with the corresponding $\lambda_{\max}=\lambda_0=1.06$, and reaches its maximum value of $\sqrt{2} -1$ (up to the set accuracy of numerical evaluation), which is the maximum attainable magic from a qubit confined to an equatorial plane of the Bloch sphere~\cite{howard,heinrich}. In figure~\ref{XY_plot}(e), we show that the globally optimal magic in the parameter space $(\lambda, \gamma)$ is created when the ground state is factorized~\cite{AdessoPRL2008, AdessoPRB2009, AmicoEPL2011}. This is a particularly remarkable result, since the single-qubit factorized ground state (FGS) is pure, the ground state for this parameter value is demonstrably a pure $H$-state. A large number of $H$-state qubits are therefore obtainable at this point in the phase diagram which we denote as $(\lambda_0,\gamma_0)$.  

We emphasize here that the above quantum many-body system provides excellent raw materials, namely, pure unencoded magic states, for the magic state injection based paradigm of quantum computation. We do not address the question of overheads in a realistic magic state factory~\cite{campbell, bravyi}. While it is possible to create such unencoded magic qubit through, for example, the non-interacting paramagnetic Hamiltonian $\sum_{j} -J \left(\sigma^{x}_{j} + \sigma^{z}_{j}\right)$, the claim of this work is the demonstration that such unencoded pure magic states can be obtained in a more realistic interacting many-body system by focussing on a point in the phase diagram, namely the FGS point, which was previously not strongly considered from a practical standpoint due to lack of other established operational resources like entanglement. We check the stability of our result against small imperfections in the tuning of the Hamiltonian by computing the single-qubit purity and its fidelity with $H$-state for small deviations from $(\lambda_0,\gamma_0)$. Even a relatively large $\sim\pm 2\%$ error in tuning the external magnetic field or $ \sim \pm 10 \%$ error in  tuning the anisotropy results in the loss of fidelity of less than $\sim 0.1 \%$.

\subsubsection{Finite size scaling}

\begin{figure}[t]
\centering
  \begin{tabular}{cc}
\includegraphics[width=0.35\linewidth]{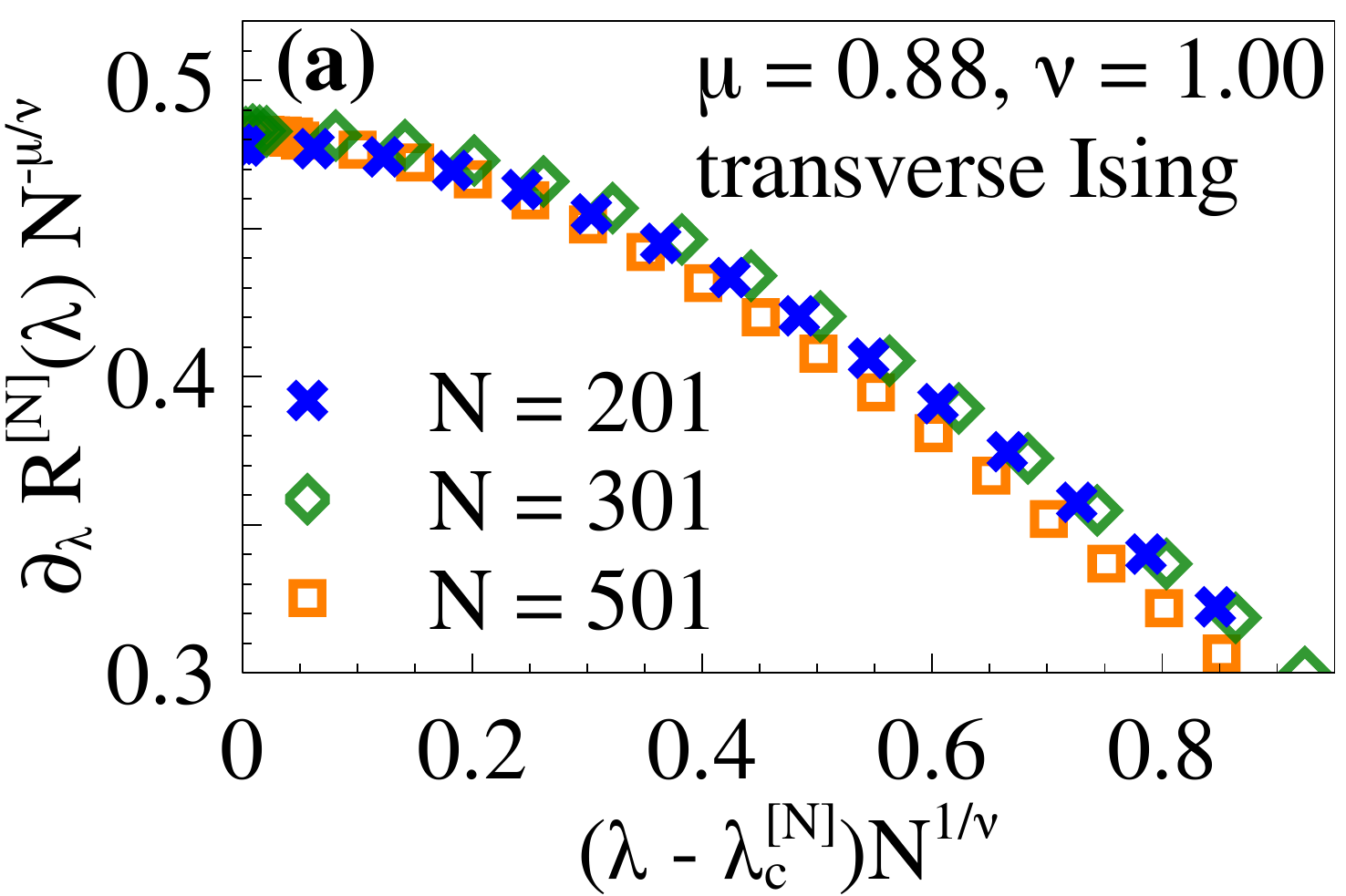}&
\includegraphics[width=0.35\linewidth]{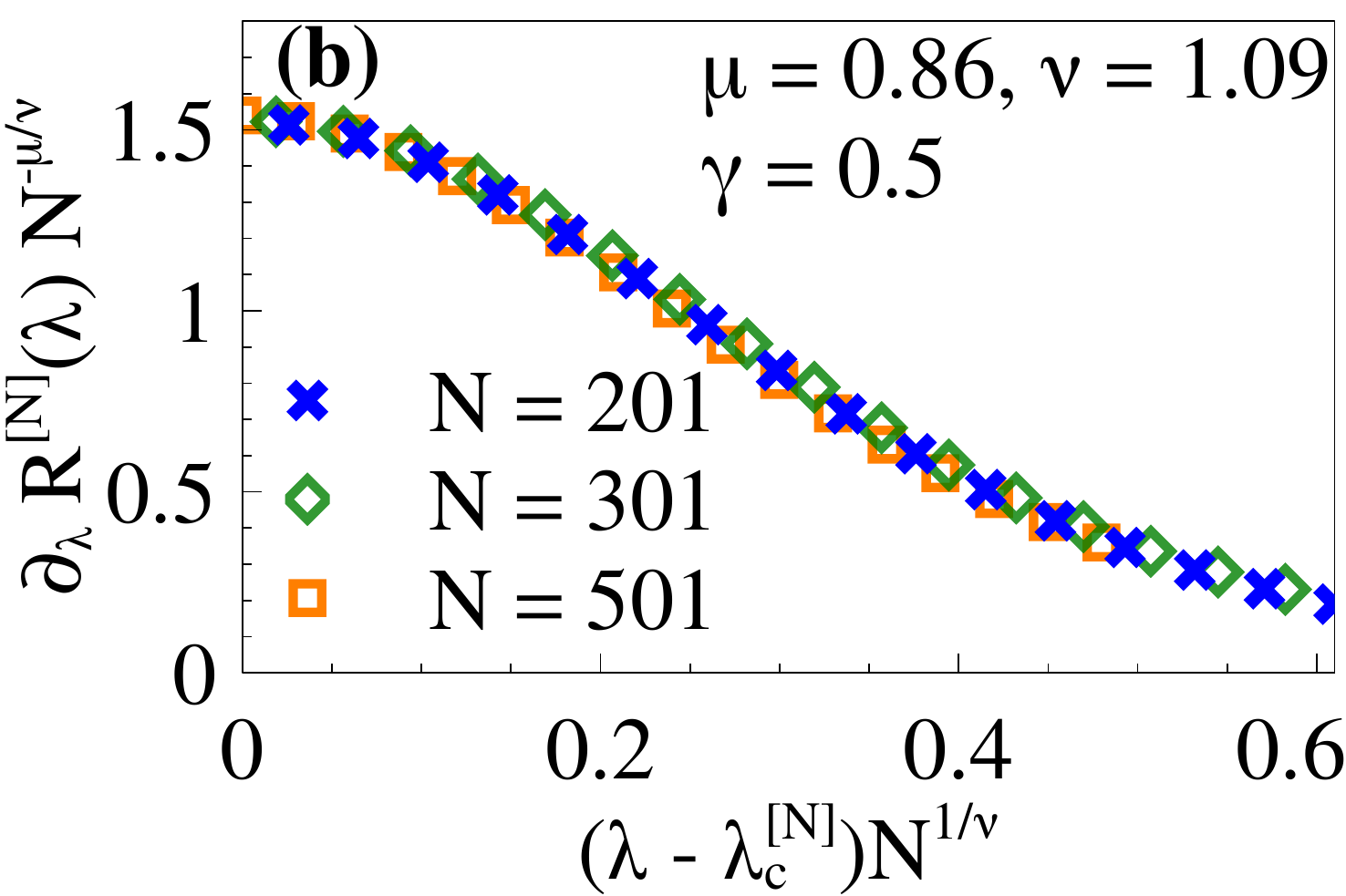}
  \end{tabular}
\caption{\textbf{Finite size scaling}. Finite size scaling of derivative of RoM for: (a) transverse Ising and (b) transverse XY chain with $\gamma=0.5$.}
\label{FSS_plot}
\end{figure}

In practice, all systems are finite, for which analytic results are difficult to obtain, since the approach of finding the magnetization $\langle \sigma^x \rangle$ from the limiting case of the corresponding two-site correlation functions fails. We use numerical methods based on density matrix renormalization group (DMRG) technique~\cite{White1992,Pirvu2010,Schollwock2011} using the ITensor Library~\cite{ITensor}, with bond dimension $300$, and magnitude of symmetry-breaking field $\sim 10^{-8}~h$, and consider the central site to minimize the boundary effects. We specifically perform the finite size scaling analysis to extract the critical exponents, some of which have been directly calculated in the previous sections. Inspired by Eq.~\eqref{derv}, we suggest a finite size ansatz~\cite{Barber1983} for the derivative of magic as $\partial_{\lambda}R_{\gamma}^{[N]} (\lambda) \sim N^{ \mu/ \nu} f\left(N^{1/\nu} (\lambda - \lambda_c^{[N]})\right)$, where, $f(\cdot)$ is an arbitrary function and $\lambda_c^{[N]}$ is the finite size critical point at which the derivative of the order parameter $\langle \sigma^x\rangle$ peaks. In Figs.~\ref{FSS_plot}(a)-(b), we plot $\partial_{\lambda}R_{\gamma}^{[N]} (\lambda) N^{ - \mu/ \nu} $ as a function of $N^{1/\nu} (\lambda - \lambda_c^{[N]})$ for various system sizes for the transverse Ising case and for $\gamma = 0.5$, respectively. By tuning $\mu$ and $\nu$, we collapse the curves corresponding to such systems. The best collapse is achieved with $\mu {=} 0.88$, $\nu {=} 1.00$ for  the transverse Ising case, and with $\mu {=} 0.86$, $\nu {=} 1.09$ for the case when $\gamma {=} 0.5$. In both cases, the exponents are quite close to the value of the scaling exponent $\mu$ extracted from the infinite chain behaviour, as demonstrated in figure~\ref{XY_plot}(c).

\subsection{Two-qubit magic analysis}

\begin{figure}[t]
    \centering
    \begin{tabular}{ccc}
      \includegraphics[width = 0.3 \linewidth]{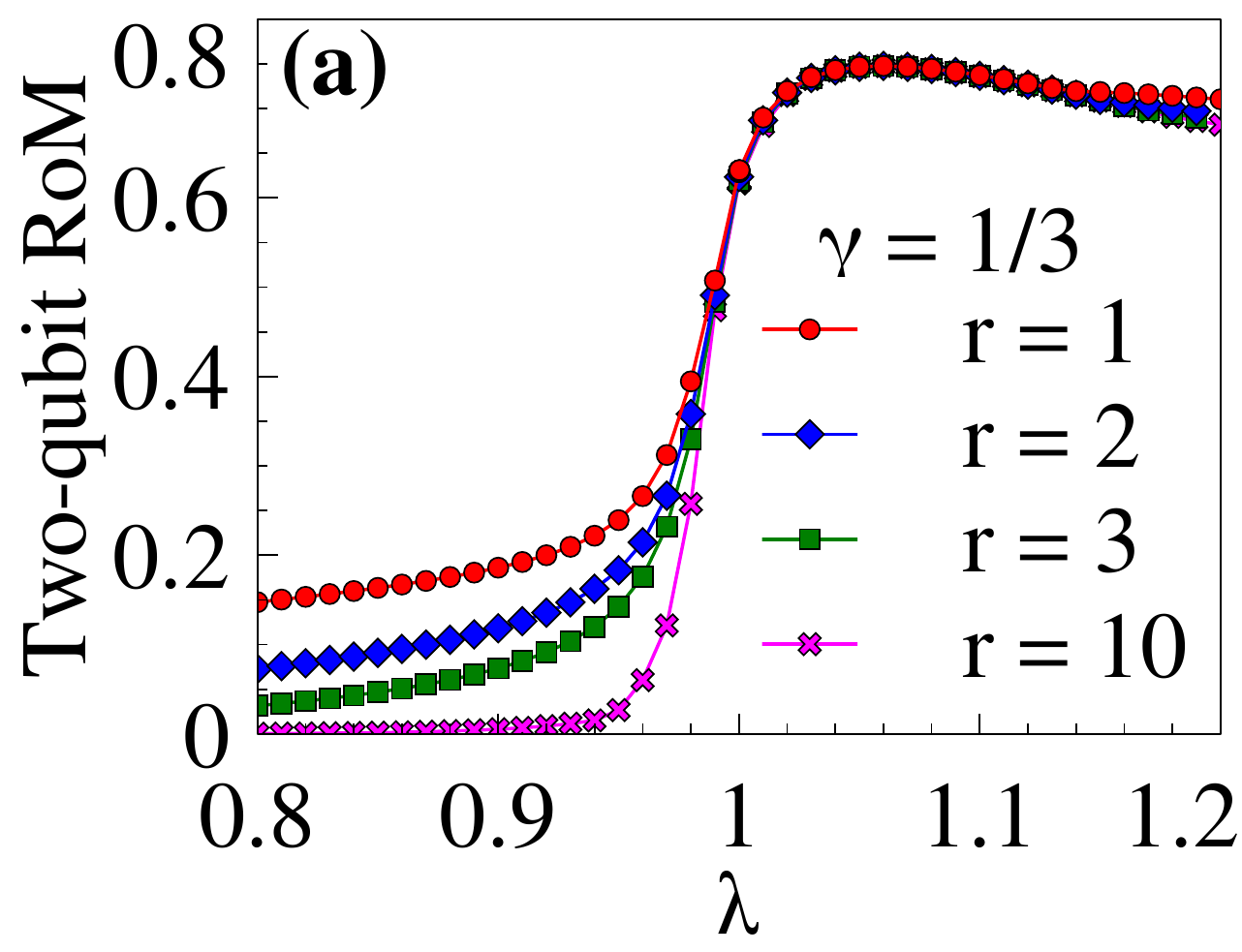} & 
      \includegraphics[width = 0.3 \linewidth]{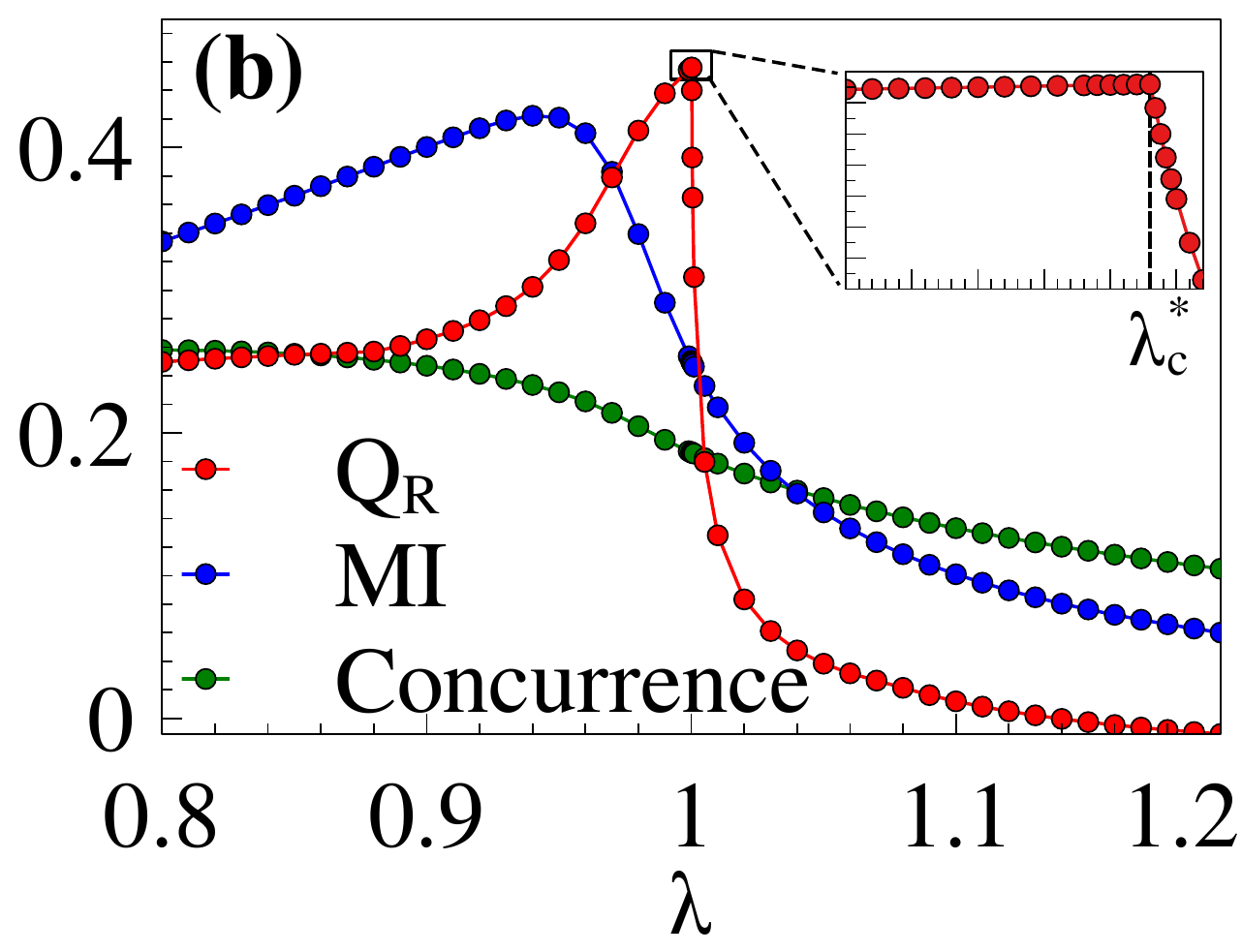} & 
      \includegraphics[width = 0.3 \linewidth]{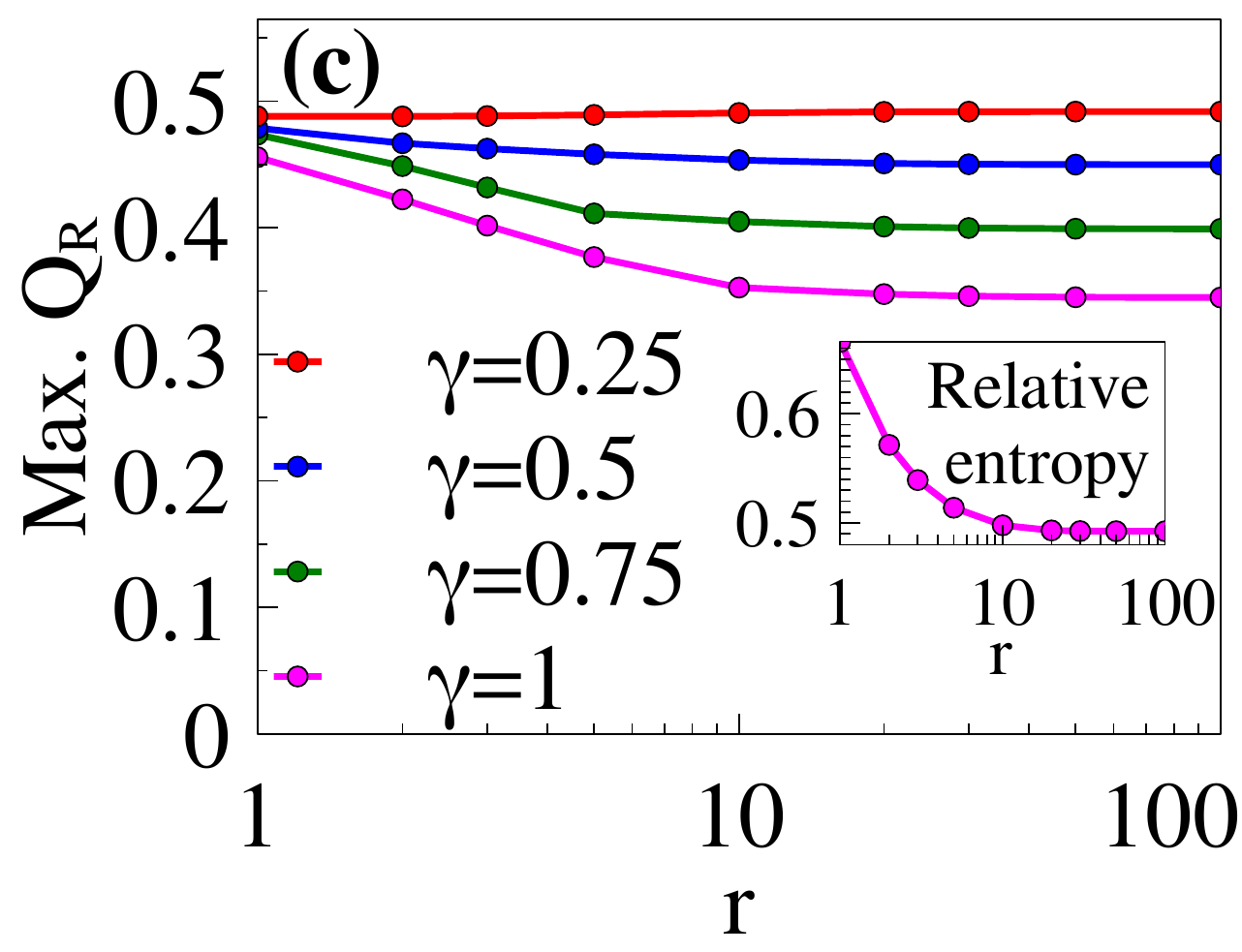}  
    \end{tabular}
    \caption{\textbf{Global magic for two qubits}. (a) RoM for two qubits at different distances in the symmetry-broken ground state for the transverse XY chain with $\gamma=1/3$. (b) Comparison of behavior of global magic $Q_R$, mutual information (MI), and concurrence, for the nearest neighbour bipartite state of an Ising chain ($\gamma = 1$). The inset shows that the global magic attains a maximum exactly at MPP. (c) The maximum value of global magic $Q_R$ at MPP stays decreases very slowly with increases in inter-site distance, with larger rate of decrease observed for more anisotropic spin chains. The inset shows similar trend displayed by the relative entropy between the reduced two-qubit state and product state of two single-qubits for the transverse Ising case.}
    \label{QR_plot}
\end{figure}

We now go on to analyse the RoM for two qubits at arbitrary distance $r$ in the symmetry-broken ground state. Unfortunately, computing the two-qubit reduced density matrix requires the knowledge of $\langle \sigma_0^{x} \sigma_r^{z} \rangle$, which is analytically intractable. Therefore, we again resort to numerical techniques based on infinite DMRG algorithm~\cite{McCulloch2008}. The two-site reduced density matrices are obtained using the ITensor Library~\cite{ITensor}, with the same  maximum bond dimension as before and a small magnetic field in the $x$-direction to break the symmetry.
The two-qubit RoM is computed by optimizing Eq.~\eqref{Magic_def}, using the convex program solver CVX~\cite{cvx1,cvx2}. As shown in figure~\ref{QR_plot}(a), the two-qubit RoM is present for arbitrary distances between the qubits, contrary to the short-ranged behaviour of two-qubit entanglement. It increases sharply as one goes from the disordered to the ordered phase. This steepness of the growth at criticality increases as the distance between the two qubits increases, and should reproduce similar qualitative behaviour as displayed by the single qubit in the limit of infinite distance. Expectedly, the maximum two-qubit RoM value of $(3\sqrt{2}-2)/3$~\cite{howard} is attained at FGS point $(\lambda_0,\gamma_0)$, as shown in figure~\ref{QR_plot}(a).

At this point, a natural question to ask is whether the two-qubit RoM reveals anything about the correlations in the symmetry broken ground state. For single-system quantum properties like coherence or entropy, a possible measure of correlation for a bipartite quantum system can be put forward as the difference between the magnitude of that property for the bipartite state, and that for the uncorrelated state which is a product of reduced individual states of the bipartite system. To wit, if $\mathfrak{C}(\rho_{12})$ is the value of such a property for a bipartite quantum state $\rho_{12}$, then the following quantity $\mathfrak{C}_{\text{global}}$ may tell us about the  correlation in the bipartite state
\begin{align}
\mathfrak{C}_{\text{global}} = \mathfrak{C}(\rho_{12}) - \mathfrak{C}(\rho_1 \otimes \rho_2)
\,.
\end{align}
If $\mathfrak{C}$ is the quantum coherence, then this quantity is known in literature as the global coherence~\cite{Radhakrishnan2016}. If $\mathfrak{C}$ is the von-Neumann entropy of a state, then this quantity is simply the mutual information. In our analysis, we assume $\mathfrak{C}$ to be the log-robustness of magic~\cite{campbell, goursanders}, that is, $\mathfrak{C} = \log (1+R) $, and concentrate on the global part of the magic defined on a two-qubit state $\rho_{12}$ as 
\begin{equation}
    Q_R = \log (1+R(\rho_{12})) - \log(1+R(\rho_1 \otimes \rho_2))
\end{equation}

This quantity is certainly nonzero for a bipartite correlated state, and its invariance under local Clifford unitaries can be easily shown~\cite{njp_magic}. Plotting this quantity in figure~\ref{QR_plot}(b) reveals several interesting features. Firstly, we know that any quantum system near criticality is scale invariant, and is accompanied by presence of long range correlations. For entanglement, its derivative diverges at the criticality, but the absolute magnitude does not attain an extremum. Moreover, since bipartite entanglement does not persist beyond the next nearest neighbour in this model, it cannot be used when searching for presence of long range quantum correlations~\cite{Osborne2002}. Discord and similar quantities also have a diverging derivative at the critical point and not an extremum~\cite{Dillenschneider2008}. However, the global magic $Q_R$ attains its maxima very close to the critical point, and in fact the point at which the maxima is obtained is exactly the magic pseudocritical point (MPP), irrespective of the inter-site distance $r$. This is in stark contrast with the case for other quantum correlation measures like entanglement, or quantum discord, which exhibit maxima for different values of the order parameter depending upon the inter-site distance~\cite{Osborne2002, Sarandy2009, Krutitsky2017}. This endows the MPP with another physical property. It is also interesting to note that at the MPP, the global magic shows a non-analyticity in the sense that the derivatives from the left and right do not match. From figure~\ref{QR_plot}(c), we also observe another interesting fact, that is, the value of global magic $Q_R$ at MPP in particular, and near criticality in general, is a slowly varying function of the distance between the qubits. This supports our claim that the global magic $Q_R$ captures long range correlations in critical quantum systems very well. In the scenario of considering  pairwise qubits, a particular manifestation of the long range correlation present at criticality is observed when considering the geometric distance between the correlated two-qubit reduced state and the uncorrelated product state composed of local single-qubit density matrices. This distance, quantified in terms of relative entropy (or, fidelity), does not decay sharply in the spin chain at criticality, which can be seen in the inset of figure~\ref{QR_plot}(c) for the Ising chain. However, more familiar measures of correlation fail to capture this feature, as may be observed in figure~\ref{QR_plot}(b) for the case of mutual information (MI), entanglement, or discord. On the contrary, the global magic $Q_R$ successfully captures the long range correlation present among qubits in the critical transverse field XY model. It is beyond the scope of the present investigation to delineate the contributions from quantum or classical nature of the correlations.

\section{Magic at thermal equilibrium}
\label{secIV}

After studying the behaviour of RoM in the symmetry-broken ground state of the spin chain, we now look at the features appearing for states in thermal equilibrium. These states now retain the phase-flip symmetry, since the presence of thermal fluctuations allow degenerate ground states to eventually become equally populated. RoM for the single qubit would be always zero as $\langle \sigma^x \rangle = 0$. Therefore, in this section we will only focus on two-qubit RoM. 

\subsection{Thermal ground state}

\begin{figure}[t]
\centering
  \begin{tabular}{cc}
    \includegraphics[width=0.35\linewidth]{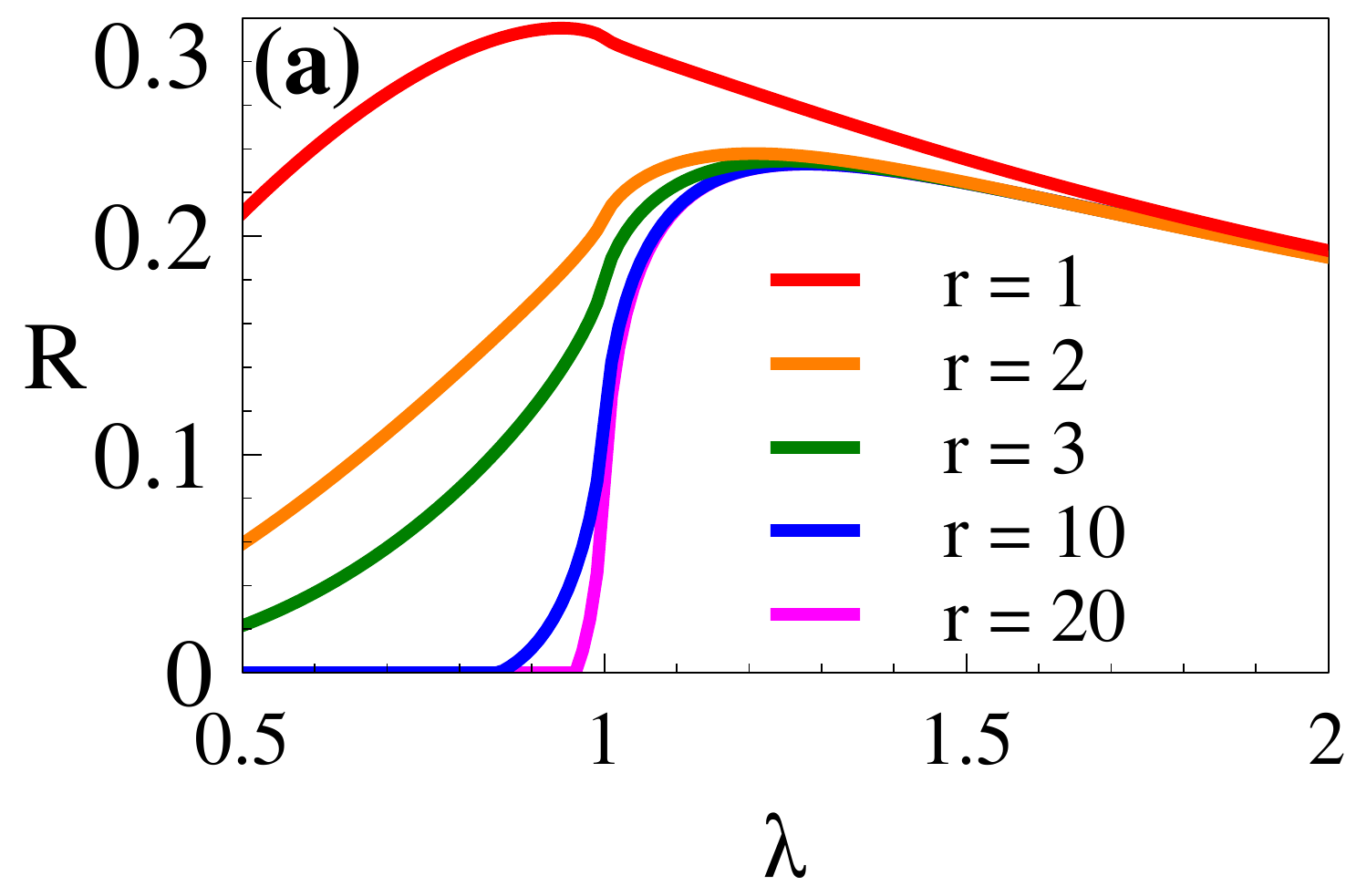} 
    \includegraphics[width=0.35\linewidth]{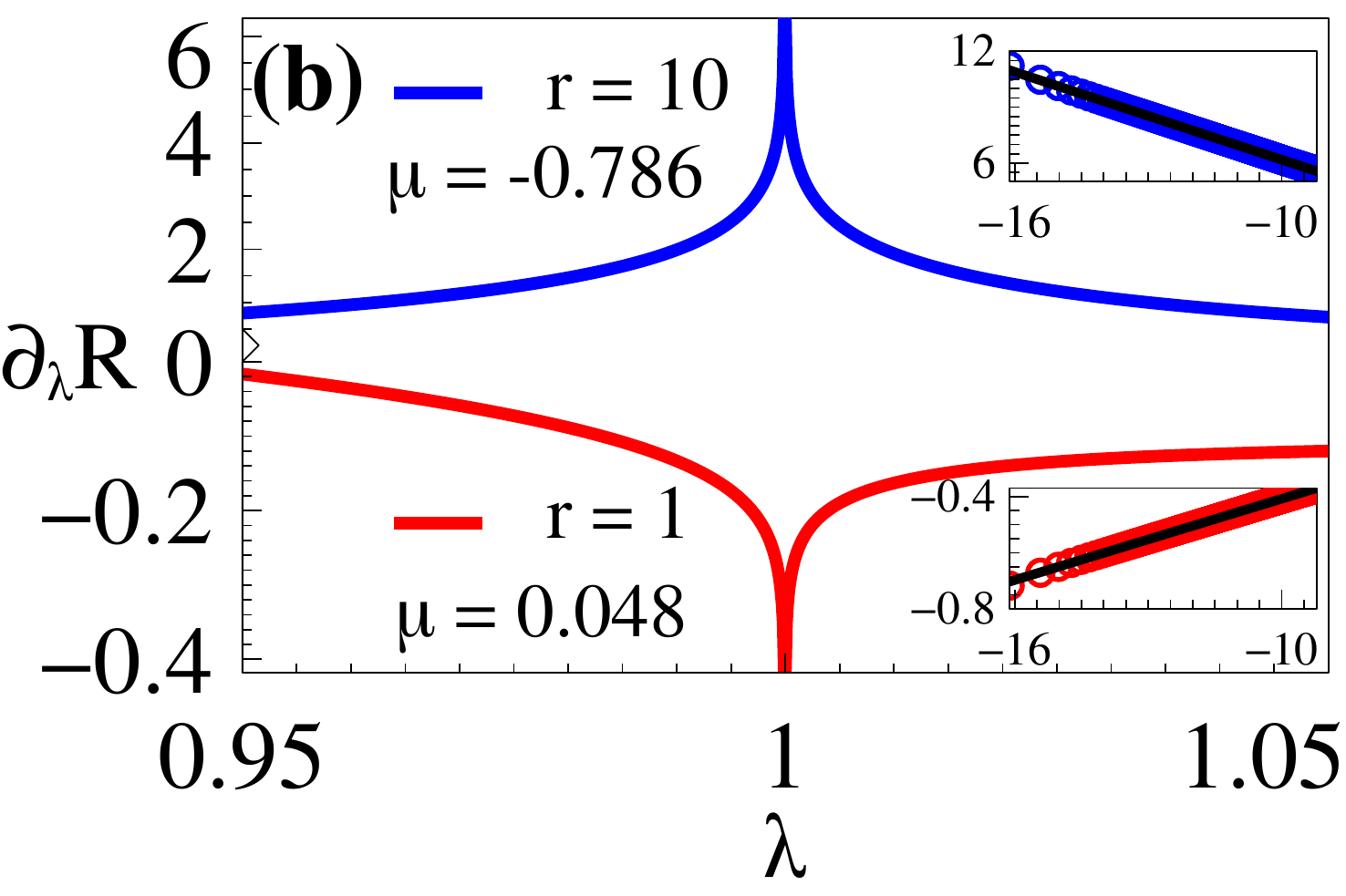} \\
    \includegraphics[width=0.35\linewidth]{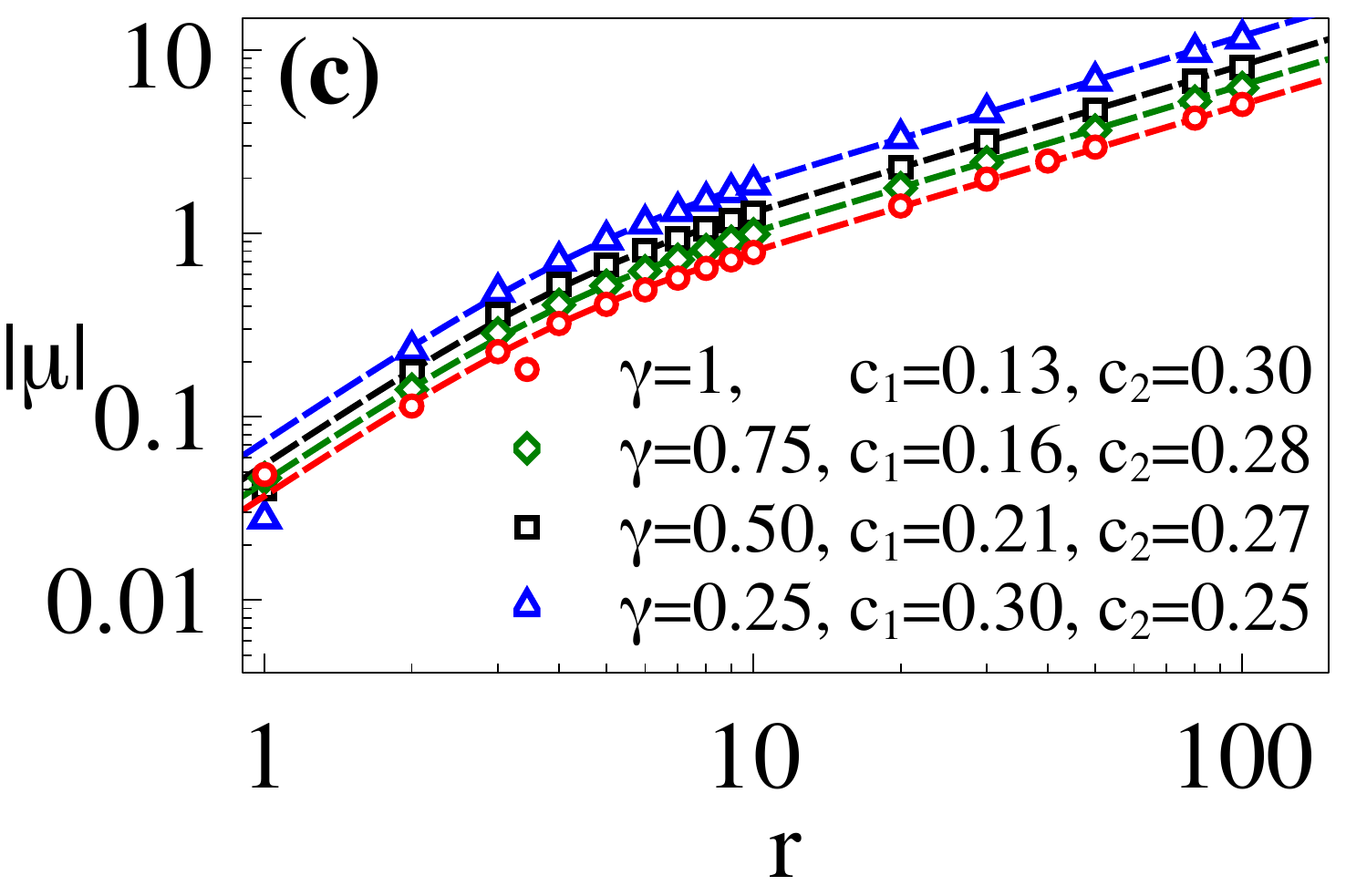} 
    \includegraphics[width=0.35\linewidth]{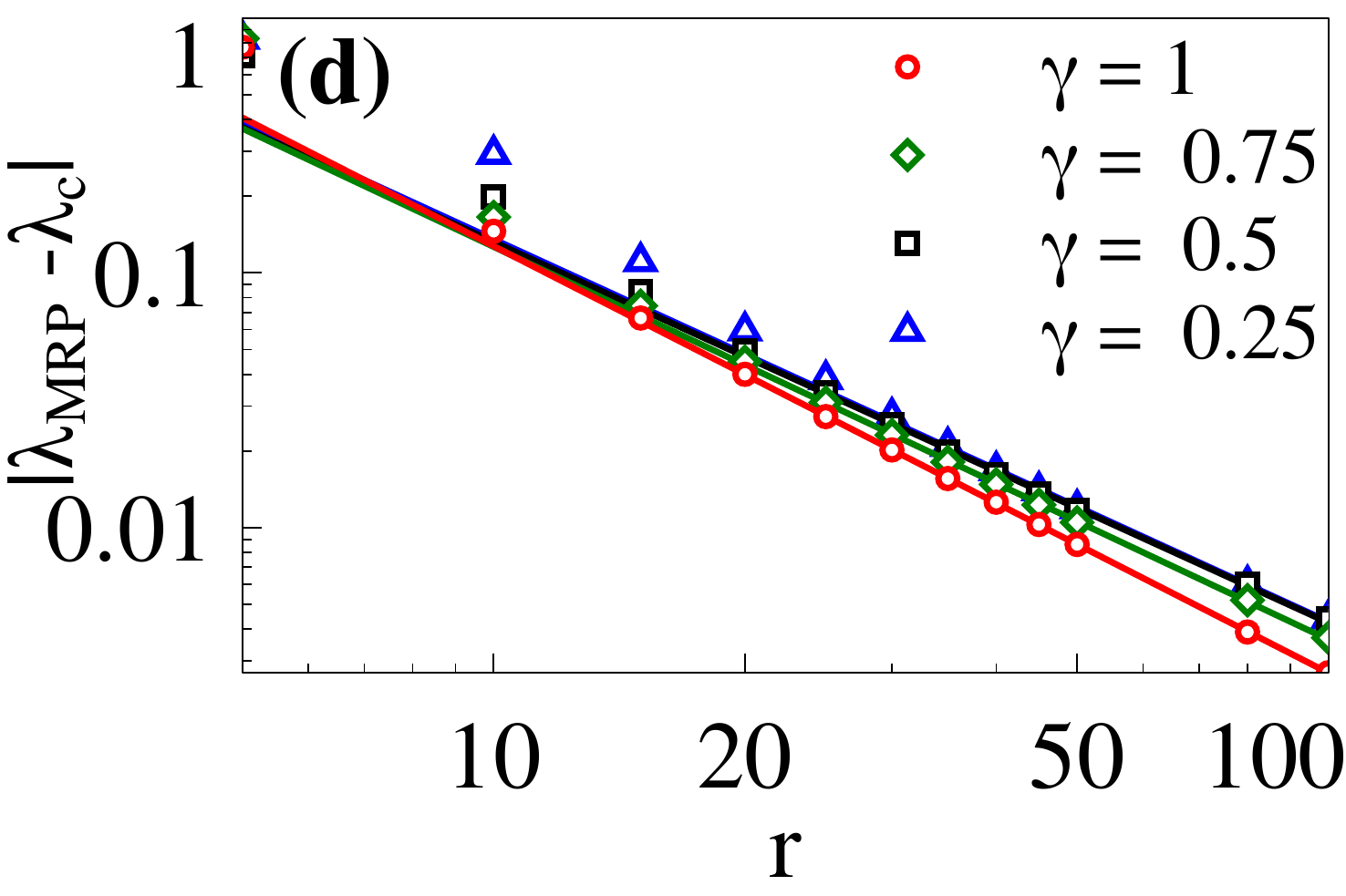}
  \end{tabular}  
\caption{\textbf{Magic in thermal ground state}. (a) RoM and MRP for TI Model at $T = 0$. (b) Logarithmic divergence of derivative of RoM for TI model for  distances $r=1$ and $r=10$. Insets show the scaling of derivative of RoM which is plotted against $\ln |\lambda-\lambda_c|$. (c) Scaling of scaling exponents with distance according to the heuristic fit $|\mu| \approx c_1 \tanh{(c_2 r)} r^{0.8}$. (d) Scaling for MRP with distance.}
\label{T=0_plot}
\end{figure}

We start with the zero-temperature limit where the density matrix is the pure ground state in absence of degeneracy, and an equal mixture of the two degenerate ground states otherwise. Taking advantage of the global phase-flip symmetry and the additional symmetries of the problem as well, namely, the translational invariance and the reality of the Hamiltonian, one can write the reduced density matrix for the sites $i$ and $i+r$ as,
\begin{align}
\rho _r= \frac{1}{4} \left( \mathds{1} + \langle \sigma^{z} \rangle \left( \sigma_{i}^{z} + \sigma_{i+r}^{z} \right) + \displaystyle\sum_{\alpha=1}^3\langle \sigma^{\alpha}_{i} \sigma^{\alpha}_{i+r} \rangle \sigma_{i}^{\alpha} \otimes \sigma^{\alpha}_{i+r} \right)  
\,,  
\end{align}
Using the expressions for the single- and two-site correlations from section~\ref{secII} we compute the density operator, and subsequently determine the RoM $R(\gamma,\lambda,r)$ for two qubits at a distance $r$ by optimizing Eq.~\eqref{Magic_def} as before. 

In figure~\ref{T=0_plot}(a) we show result for the transverse Ising case i.e., $\gamma=1$. A closer look at this figure reveals a diverging nature of $\partial_{\lambda} R$ near criticality which is studied next and is shown in figure~\ref{T=0_plot}(b). Indeed, we find a logarithmic divergence,
\begin{align}
\partial_{\lambda} R \approx \mu \ln|\lambda-\lambda_c| + \text{const.}
\,.
\end{align}
The scaling exponent $\mu$ is positive for $r=1$ and negative for $r>1$, as is shown with the help of the numerical fits in the inset of figure~\ref{T=0_plot}(b) for $r=1$ and $r=10$. 

We now calculate the two-qubit RoM $R$ in the $\gamma-\lambda$ plane for various distances. The results are qualitatively similar to that for the transverse Ising case. The peak of $R$ is usually reached in the ordered phase, except for the cases of $\gamma > 0.56$ where the peak for only $r=1$ is found in the disordered phase. The divergence at criticality is analysed again, and absolute value of the exponent $|\mu|$ is found to increasing with $r$ for a given $\gamma$. Moreover, we found a fit for such behaviour: $|\mu| \approx c_1 \tanh{(c_2 r)} r^{0.8}$, where $c_1, c_2$ are constants that depend on the anisotropy parameter $\gamma$. This is depicted in figure~\ref{T=0_plot}(c). Another interesting kind of scaling appears if we consider the magic rising point ($\lambda_{\text{MRP}}$) of the system which we define as the lowest value of $\lambda$ for which $R$ is non-zero. From figure~\ref{T=0_plot}(a), it is evident that MRP increases with increasing $r$. In figure~\ref{T=0_plot}(d) we show that the distance between $\lambda_{\text{MRP}}$ from the critical point goes as $\ln r$, for large values of $r$. We note in passing that the two-qubit discord also shows persistence over long distance for the transverse XY model. However, for quantum discord, the first derivative does not show a pronounced peak at criticality, and a logarithmic divergence at criticality is observed only for the second derivative with respect to the order parameter $\lambda$~\cite{Sarandy2009}. In this respect, two qubit magic for arbitrary distances behaves rather similarly to nearest neighbour entanglement than discord, in the sense that the first derivative of the two qubit magic does show a logarithmic divergence at criticality~\cite{Osterloh2002}.

\subsection{Thermal behaviour at criticality}

So far we have focused our attention on the quantum critical point at $T = 0$. In the finite temperature scenario, thermal noises would dominate at higher temperatures and a classical phase transition would be observed in the system at some critical temperature. However, in the intermediate regime, both quantum fluctuations as well as thermal fluctuations are significant, and hence instead of a single critical point at $T = 0$, the existence of a quantum critical region is posited~\cite{Sachdev1999}. The effect of quantum fluctuations is to be captured by the deviation of the Hamiltonian parameter $\lambda$ from unity, while the strength of the thermal fluctuations is indicated by the temperature $T$. The quantum critical region in the $(T, \lambda)$ parameter-space is supposed to fan out conically from the single quantum critical point at $ (T, \lambda ) = (0,1)$. From the perspective of pairwise quantum entanglement, this effect was indeed observed in~\cite{Amico2007, Frerot2019}. In the final part of the present work, we demonstrate that two-qubit RoM in the spin chain can also detect the existence of the quantum critical region, retaining largely the same qualitative features. As before, the fact that pairwise magic persists for long distances, unlike entanglement, which is absent in qubit pairs more than two sites away, may present usefulness in a less constrained characterization of the quantum critical region. 

\begin{figure}[t]
\centering
  \begin{tabular}{cc}
	\includegraphics[width = 0.35 \linewidth]{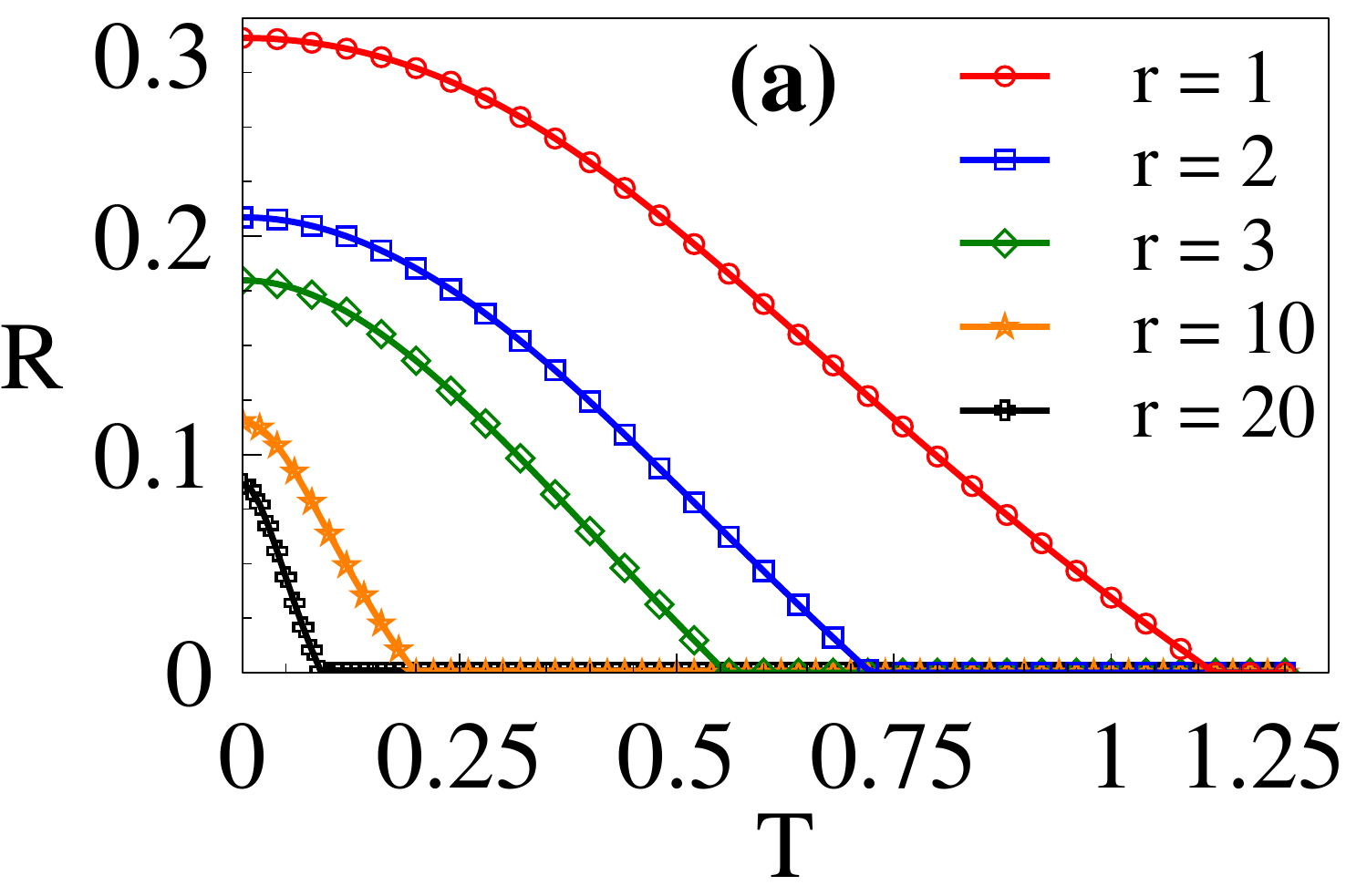} 
	\includegraphics[width = 0.35 \linewidth]{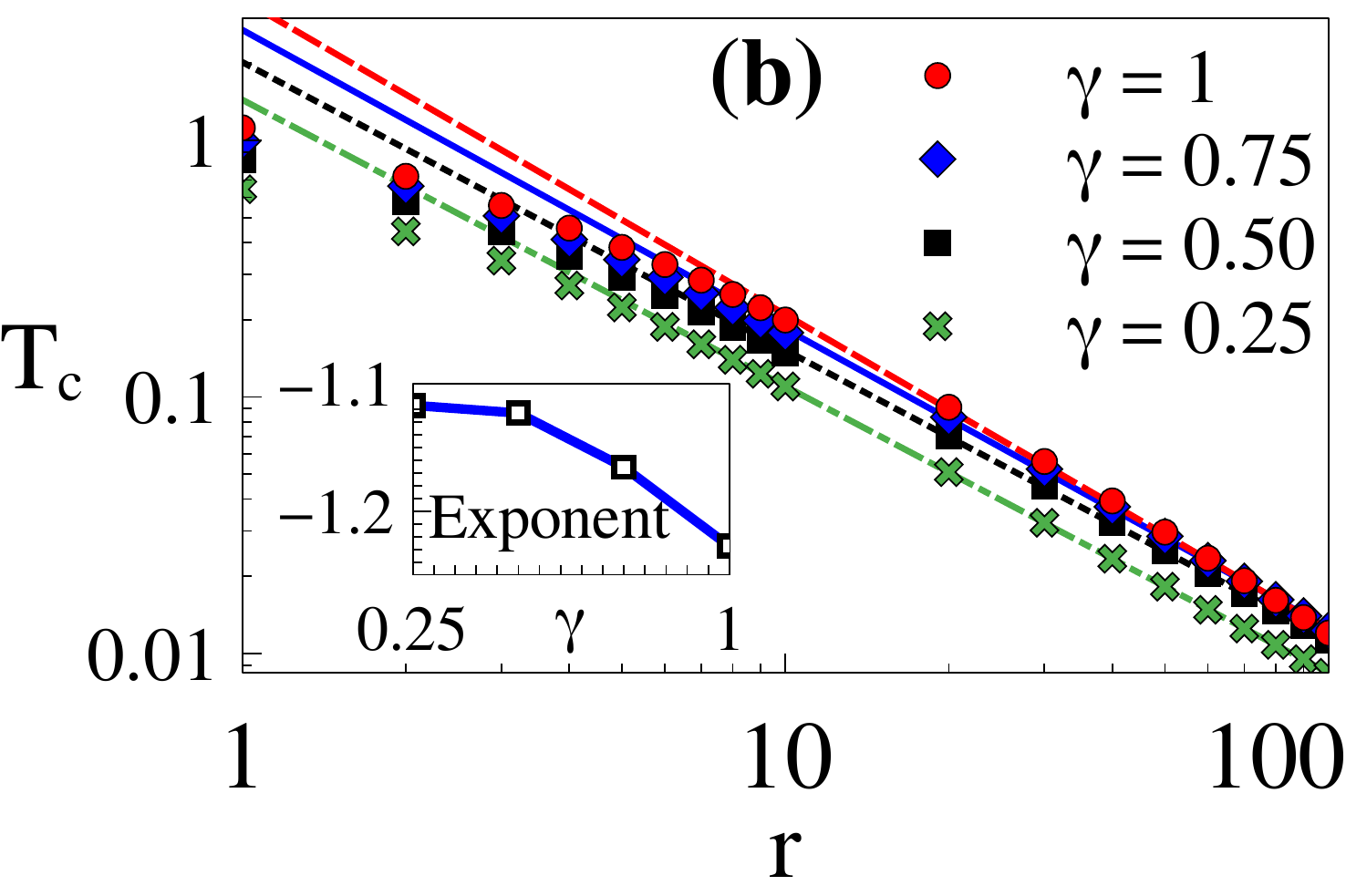} \\
	\includegraphics[width = 0.35 \linewidth]{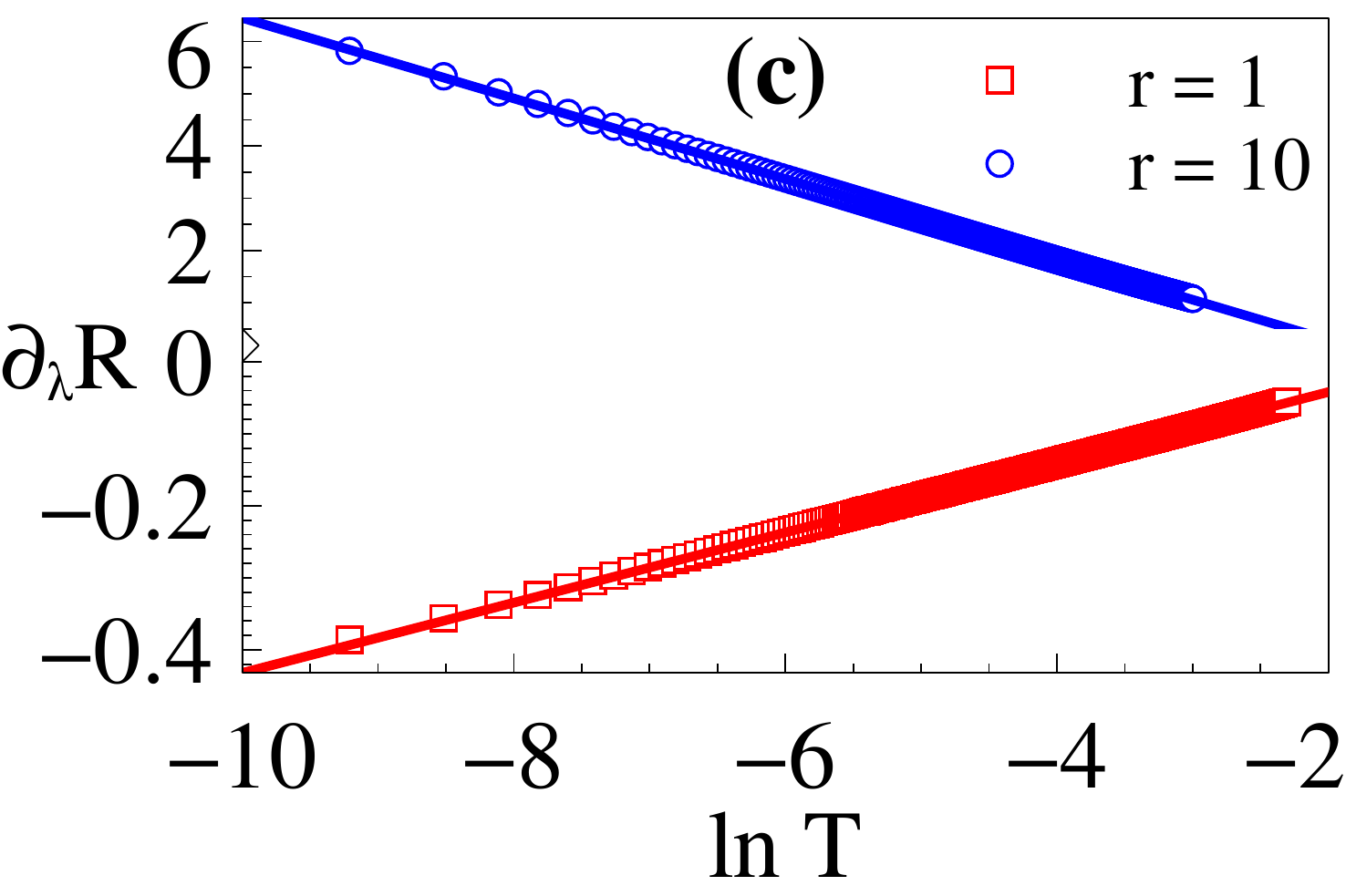} 
	\includegraphics[width = 0.35 \linewidth]{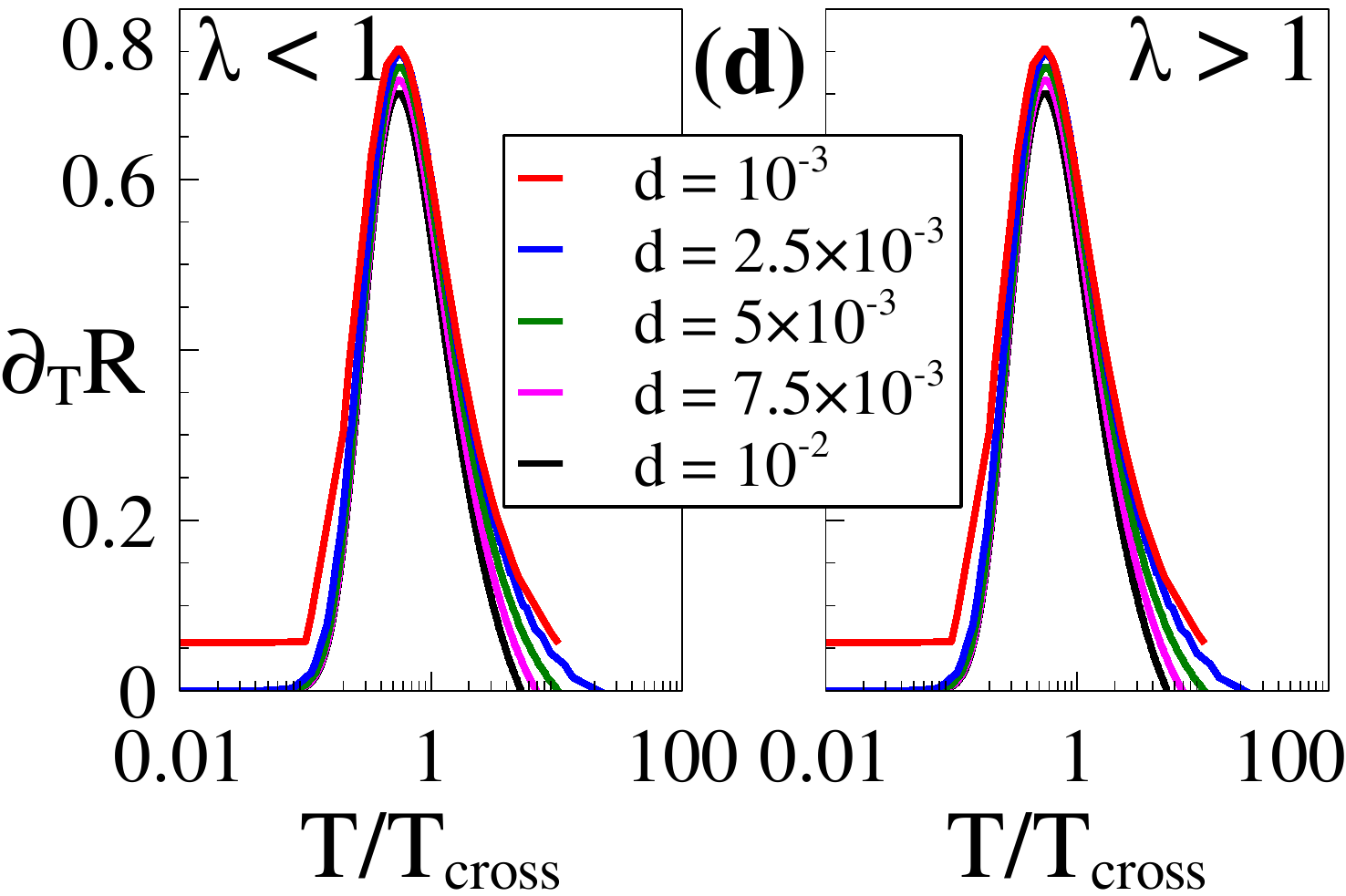}
  \end{tabular}  
\caption{\textbf{Magic in thermal equilibrium}. (a) Sudden death of RoM at criticality with increasing temperature for the transverse Ising chain. (b) Scaling of sudden death temperature of RoM at criticality. (c) Logarithmic divergence for RoM at criticality of the form $\partial_{\lambda} R = c + \xi \ln T$ for the transverse Ising case. (d) Collapse of thermal fluctuations captured through the derivative of magic in transverse Ising chain with respect to temperature at the crossover temperature $T_{\text{cross}}$ irrespective of the Euclidean distance $d$ from the quantum critical point in the $\lambda$-$T$ plane for inter-site separation $r=10$. }
\label{Thermal_plot}
\end{figure}

Let us first concentrate on the effect of thermal noise on magic content at the critical point $\lambda = \lambda_{c} = 1$. In addition to the significance in characterizing the nature of thermal fluctuations for finite temperatures, this is obviously important from the operational standpoint as well, provided quantum computing gates utilizing these qubits as magic state ancillae are being considered. We demonstrate in figure~\ref{Thermal_plot}(a) that the magic content of the two qubits decreases monotonically with increasing temperature, and vanishes at a certain finite temperature. The situation is roughly similar to that of nearest neighbour entanglement except for one detail. At very low temperature, the entangled ground state may get mixed with an even more entangled excited state, thus leading to an increase in entanglement vis-a-vis the $T=0$ scenario, a phenomenon that is often called \emph{entanglement mixing}~\cite{Osterloh2002}. However, for the case of bipartite magic, we could not find any evidence of such \emph{magic mixing}, and the quantity of magic monotonically decreases with temperature, finally vanishing at a \emph{sudden death point} at some finite temperature. It is noticeable that the temperature at sudden death point monotonically decreases as the distance between the two qubits is increased. Similar results are obtained if more general transverse XY chains are considered.
 
The above observation is quantified in figure~\ref{Thermal_plot}(b), where we demonstrate that, especially for larger distances, the sudden death temperature $T_c$ at criticality follows a scaling law with the distance between the two qubitss $r$, expressed as 
\begin{equation}
T_c \propto r^{\kappa},
\end{equation}
where $\kappa$ is a function of the anisotropy parameter $\gamma$ of the transverse XY chain. We also demonstrate in the inset of this figure that $\kappa$ decreases slowly as the anisotropy of the transverse XY chain is increased, and its magnitude is maximum for the Ising case.

\emph{Low temperature scaling --} We have previously demonstrated that at zero temperature, the derivative of magic with respect to the order parameter $\lambda$ diverges at $\lambda = \lambda_c = 1$, signifying the quantum phase transition. It is thus a natural question to ask what happens to the quantum fluctuations, signified by the behaviour of the derivative of magic at criticality, if the temperature $T$ being considered is finite, yet lower than the sudden death temperature of magic $T_c$. We observe the following general behaviour of the derivative of magic in these cases
\begin{equation}
\partial_{\lambda}R = c + \xi \ln T,
\end{equation}
where the constants $c$ and $\xi$ depend on the anisotropy parameter $\gamma$, as well as the distance $r$ between the two qubits. This is similar to the behaviour of entanglement~\cite{Osborne2002}. Figure~\ref{Thermal_plot}(c) shows this for the transverse Ising case, although this is observable for general anisotropy parameter values.

\subsection{Identifying quantum critical region through pairwise magic}

Quantum criticality occurs at $T = 0$, which is unreachable in principle. However, even at finite temperatures, it is possible to surmise the existence of quantum criticality through a crossover phenomenon between the long-range ordered, i.e., renormalized-classical, and the disordered regimes. At low temperatures, in addition to quantum fluctuations of the order parameter $\langle \sigma^x \rangle$, the thermal noise due to temperature becomes important. For $T= 0$, there are only two regions in the phase diagram, namely the long range ordered region associated with $\lambda > 1$ and the disordered regime associated with $\lambda < 1$. However, for finite temperatures, these regimes cross over into each other through the so-called `quantum critical region', characterized by the crossover temperature $T_{\text{cross}} = |\lambda - \lambda_{c}|^{z \nu}$, where $z = \nu  = 1$ for the Ising universality class~\cite{Continentino2001}. If the temperature is much lower than $T_{\text{cross}}$, the thermal excitations, characterized by the thermal de-Broglie wavelength, are not sufficient to induce transitions from ground to excited states on their own. Hence the correlation function separates out into two contributions coming from quantum and thermal fluctuations, respectively.  On the other hand, when $T \gg T_{\text{cross}}$, the thermal wavelengths are of the same order as the spacings of excitations, and such a separation can no longer be maintained. The latter regime is the quantum critical region, where the physics is dominated by the interplay between these two fluctuations. 

\begin{figure}[t]
\centering
  \begin{tabular}{cc}
	\includegraphics[width = 0.35 \linewidth]{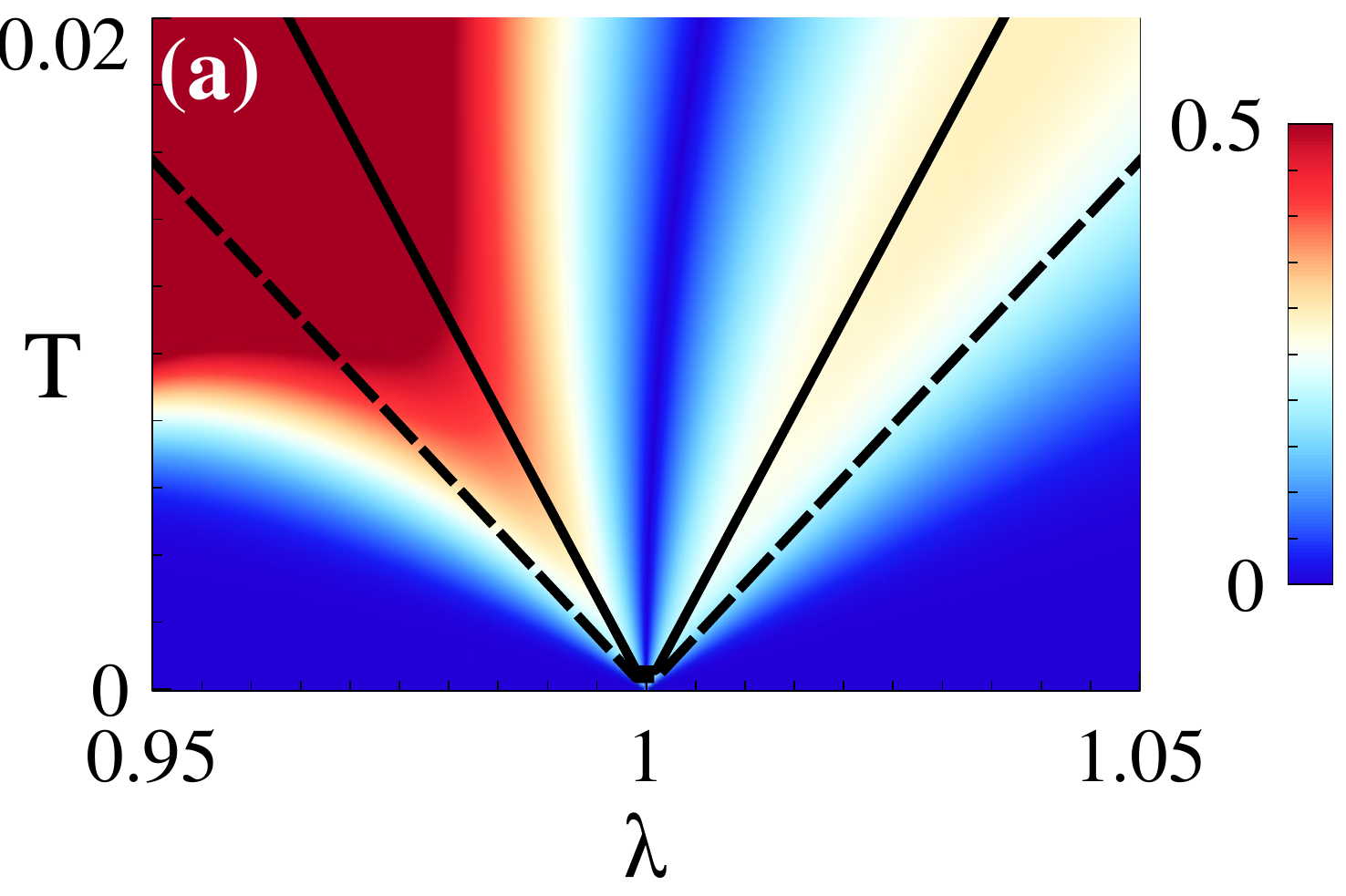} 
	\includegraphics[width = 0.35 \linewidth]{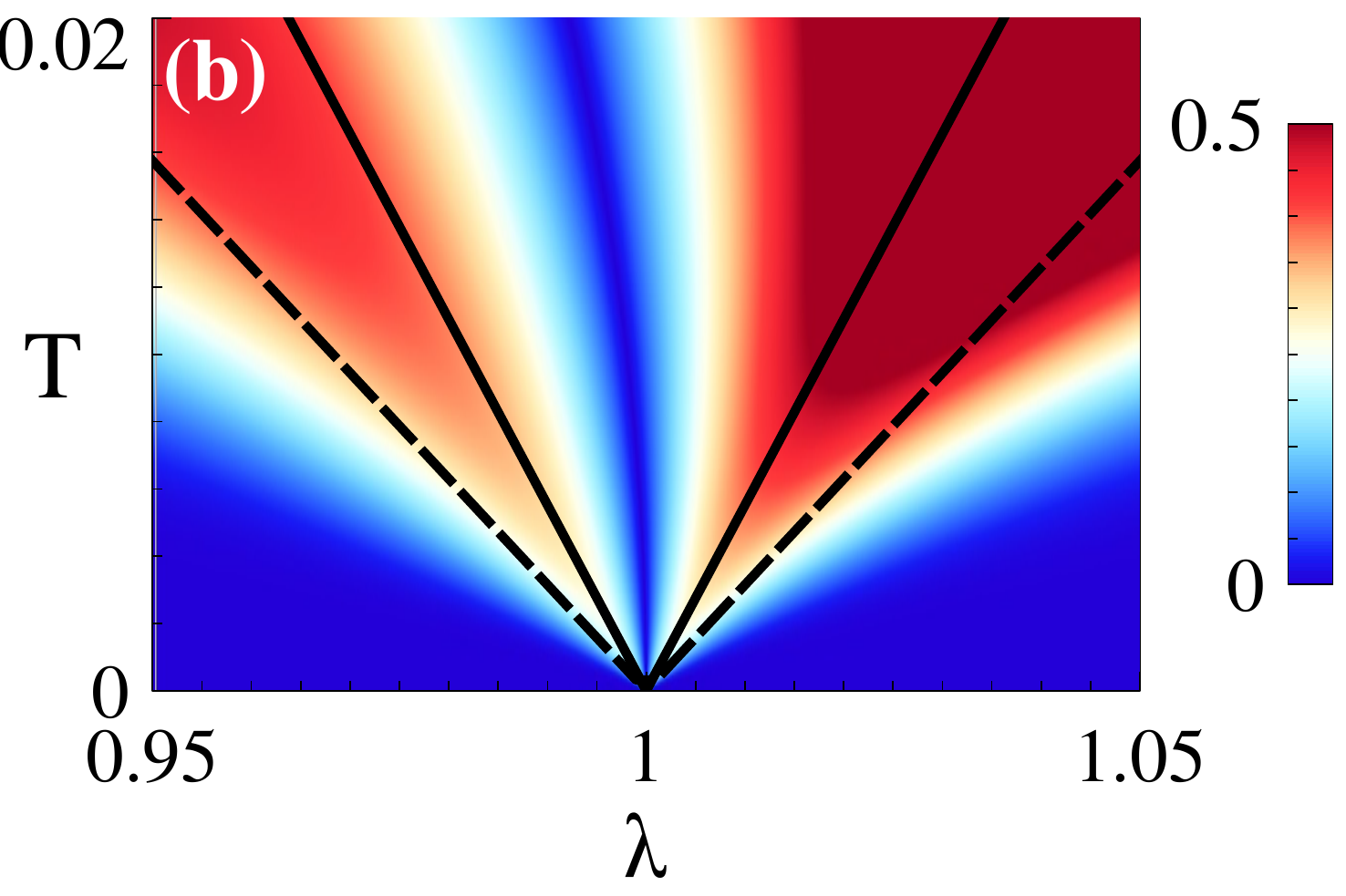}\\
	\includegraphics[width = 0.35 \linewidth]{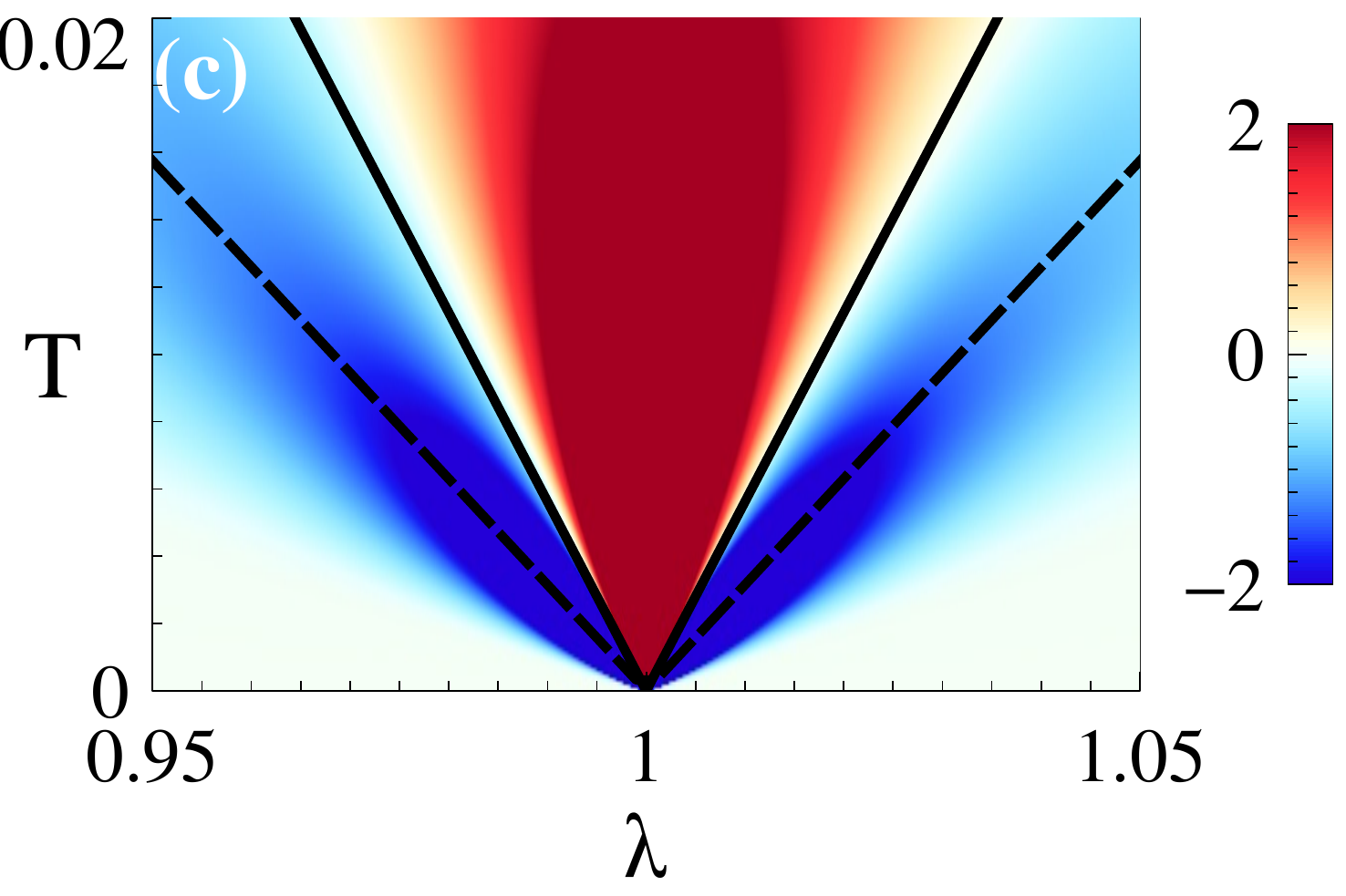} 
	\includegraphics[width = 0.35 \linewidth]{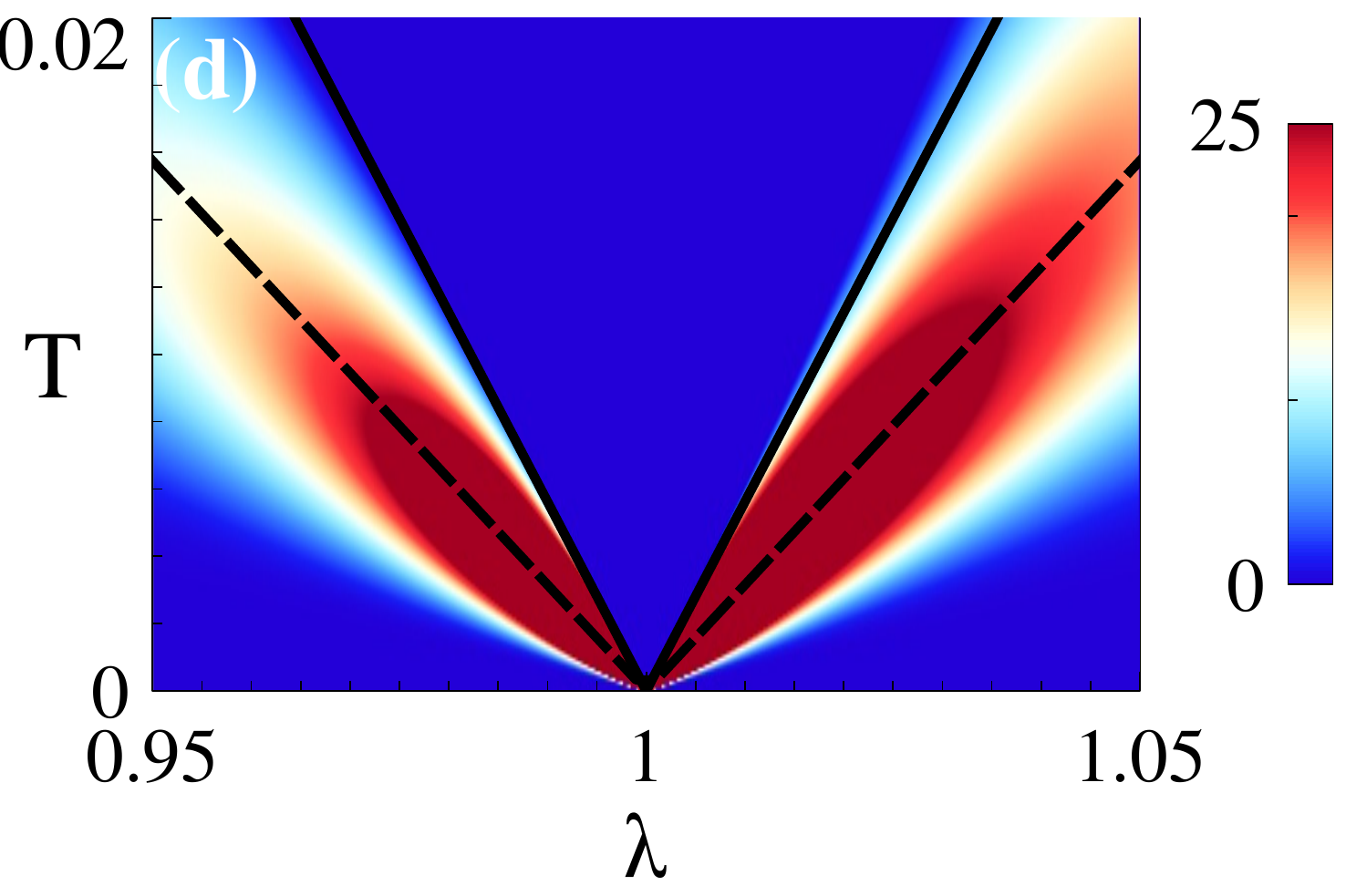}
  \end{tabular}	 
\caption{\textbf{Critical region for transverse Ising chain}. Gr{\"u}neisen parameter $|\partial_{T}R|/|\partial_{\lambda}R|$ for (a) $r = 1$, and (b) $r = 10$, and Second derivative $\partial_{T} \partial_{\lambda} R(r)$ for (c) $r = 1$, and (d) $r = 10$. Dashed line indicates the extrema of $\partial_{T} \partial_{\lambda} R(r)$, as expressed by $T_{M} = b T_{\text{cross}}$, with $b \approx 0.31$ irrespective of intersite distance, and solid line indicates $T^{*} = a T_{\text{cross}}$ ($a = 0.56$ for $r = 1$, and $a = 0.55$ for $r = 10$).  }
\label{Crossover_plot}
\end{figure}

By the theory of quantum critical phenomena~\cite{Continentino2001}, the scaling ansatz for the derivative of magic with respect to the order parameter $\lambda$ for two qubits at sites $r$-distance apart is given by 
\begin{equation}
\partial_{\lambda} R (r) \approx \ln \left[ T^{n} f_1 \left( \frac{T}{T_{\text{cross}}} \right) \right],
\end{equation}
and the derivative of magic with respect to temperature, signifying the thermal fluctuations are to obey the following scaling ansatz
\begin{equation}
\partial_{T} R(r) \approx f_2 \left( \frac{T}{T_{\text{cross}}} \right),
\end{equation}
where $f_1$ and $f_2$ are the scaling functions and the exponent $n$ is non-universal. Function $f_2$ is characterized by a maxima that collapses for all those points in the $(\lambda, T)$ phase diagram, whose distance from quantum critical point is given by $d = \sqrt{T^2 + (\lambda -\lambda_{c})^2}$. The maxima, at the crossover temperature $T^{*} = a T_{\text{cross}}$, is verified in figure~\ref{Thermal_plot}(d) to be almost independent of the distance $r$ between the two qubits in transverse Ising chain. For example, at $r=10$ (shown in figure~\ref{Thermal_plot}(d)), $a \approx 0.55$, and for $r=1$ (not shown), $a \approx 0.56$.  Therefore the magic experiences the maximal variation in thermal fluctuation at $T = T^{*}$, signifying the crossover phenomenon.

We have previously mentioned that the quantum critical region is characterised by the interplay between the thermal and the quantum fluctuations, even when $T \gg T_{\text{cross}}$. The strength of these fluctuations, characterized by the Gr{\"u}neisen parameter~\cite{Zhu2003} or the ratio of the absolute values of $\partial_{T} R(r)$, and $\partial_{\lambda} R(r)$, is depicted in figure~\ref{Crossover_plot} (a), (b). As is expected for the quantum critical region, the ratio is very small, signifying the dominance of quantum fluctuations for temperatures far in excess of the crossover temperature $T_{\text{cross}}$, which is the signature of the quantum critical region. 

Let us now concentrate on the response of the quantum fluctuations upon variations in temperature, i.e., $\partial_{T} \partial_{\lambda} R(r) $. In the quantum critical region, since $T \gg T_{\text{cross}}$, the thermal de Broglie wavelength is of the same order as the spectral gap, and the system responds to any infinitesimal change in the temperature.  We observe, similar to entanglement, that this second derivative attains an extremum along the line $T_{M} = b T_{\text{cross}}$. For a representative ($r = 10$) case, we observe in figure~\ref{Crossover_plot}(b), that for $T > T^{*}$, any reduction in the temperature $T$ leads to the increase of the effects of the criticality, while for $T < T^{*}$, the reduction in temperature leads to a net decrease in the effects of the criticality. This is along similar lines to behavior of entanglement. Another interesting aspect is the observation that as one moves away from the critical point, the quantum critical region identified by magic in nearest neighbour qubits ($r=1$), is skewed in a different direction than the critical region identified by magic in more distant neighbours owing to the opposite signs of the derivative of magic near criticality for nearest neighbour qubits. These, in conjunction, serve to capture the quantum critical region in detail. We add that although the results shown in this subsection pertain to the transverse Ising chain, they similarly hold for differing magnitudes of anisotropy for the more general transverse XY chain.  Similar to entanglement, magic is a purely quantum feature of a system, hence our result supports the claim of Ref.~\cite{Amico2007} that purely quantum effects like entanglement serves to characterize the crossover between quantum critical and renormalised classical regions better than merely classical effects.

\section{Conclusion}
\label{secV} 

In this paper, we have studied transverse field anisotropic XY model near criticality for single- and two-qubit magic content. In the physically relevant symmetry broken ground state, we quantify scaling behaviours of the single-qubit RoM and identify that the factorization point in the phase space provides perfect unencoded magic states. This endows the FGS with an operational significance in terms of presence of a resource for fault tolerant quantum computation for the transverse XY model. It will be interesting to investigate in future whether this phenomenon of FGS being associated with maximal magic content is persistent for other quantum many-body systems. Furthermore, we show that, similar to discord, the two-qubit RoM persists over long distances, which is a direct consequence of the presence of long-range correlation in this critical system. We show that the sharp maximum attained by the global component of magic is located at the exact point separating the magical and non-magical regimes irrespective of the inter-site distance, which is extremely near the critical point (at $\lambda \approx 1.00015$). These properties are in marked contrast with the smooth peak observed for entanglement and discord. In the symmetry unbroken case, which is relevant for the thermal states, scaling behaviours have been identified for two-qubit RoM in the zero temperature limit. The effect of the thermal fluctuation is shown to be destructive resulting in a sudden death of RoM. The competition between quantum and thermal fluctuations extends the quantum critical point into a critical region which can also be identified using RoM of two distant qubits.

With our understanding of the behaviour of RoM in the anisotropic spin chain, we can now compare magic with other established resources, namely, entanglement, discord, and coherence, in this system, which is summarised in Table~\ref{compare_table}. In the symmetry unbroken case, both discord~\cite{Maziero2010, Maziero2012} and magic show long range survival, in contrast to bipartite entanglement which does not survive beyond the next nearest neighbour. However, logarithmic divergence at criticality is shown by the second derivative of discord, compared to the first derivative in case of magic. The similarity of behaviour of magic with entanglement is observed in terms of thermal sudden death, which is not as pronounced for discord and coherence. It may be noted in this context that although discord does not undergo sudden death in this spin system, a sudden transition induced by environmental interaction during Markovian evolution has been reported~\cite{Auccaise2011,Cornelio2012,Paula2013}. In the symmetry broken case, any correlation including discord and entanglement is absent at the FGS, and the single site quantum coherence is finite but non-maximal. In contrast, the single site magic attains its global maximum at the FGS. 

This study can be extended for other many-body systems and there are various avenues to explore including dynamical phase transitions and open system evolution. We add in this context that in previous years, there has been a growing interest in dynamical evolution of quantum properties in spin systems, as well as quantification of magic content of quantum channels~\cite{Wang2019}. While we focus on the static properties of magic in the spin chain in this work, a thorough study of open system evolution of magic content can shed further light on how decoherence affects fault tolerant quantum architectures in realistic scenarios.

\begin{table}[t]
\centering
\begin{tabular}{|c|c|c|c|c|c|} \hline
Property & Entanglement & Discord & Coherence & Magic \\ \hline
Range of persistence ($T=0$) & 2  & Asymptotic & -- & Large \\ \hline
Sudden death at finite $T$ & Yes & No & No & Yes \\ \hline
Presence at FGS & No & No & Not maximal & Maximal \\ \hline
\end{tabular}
\caption{Comparison of resources for anisotropic XY chain with transverse field.}
\label{compare_table}
\end{table}

\ack{Discussions with Anton Buyshkikh and Sougato Bose are warmly acknowledged. SS acknowledges the support of the (Polish) National Science Center Grant No.~2016/22/E/ST2/00555. CM acknowledges doctoral fellowship from the Department of Atomic Energy, Government of India, as well as funding from INFOSYS. CM thanks University College London and the UESTC for their kind hospitality. AB acknowledges the National Key R\&D Program of China, Grant No.~2018YFA0306703. The computational works were performed using the Interdisciplinary Centre for Mathematical and Computational Modelling, University of Warsaw (ICM), under Computational Grant No.~G75-6.}

\appendix

\section{Simpler formula for RoM}
\label{Appendix A}

In the spin chain model considered in the paper, the reduced density matrix has zero magnetization along the $y$-axis, i.e., $\langle \sigma^y \rangle = 0$. Any such qubit state $\rho$ can be expressed as the (possibly non-convex) sum of four stabiliser states. We assume thst, the optimal decomposition, in the context of calculating the RoM, is of the following form
\begin{equation}
\rho = a_1 |0\rangle\langle 0 | + a_2 |1\rangle\langle 1 | + a_3 |+\rangle\langle +| + a_4 |-\rangle\langle - |.
\end{equation} 
The expectation values are $\langle \sigma^x \rangle = a_3 - a_4$, and $\langle \sigma^z \rangle = a_1 - a_2$, which are both positive . Now, If all of the $a_{i}$'s are non-negative, the state is within the stabiliser polytope. On the other hand, these quantities can not all be negative because of the positive semi-definiteness of density matrix. This leaves us with three different alternatives. 

\begin{enumerate}
\item \emph{If three of the coefficients are negative --}  We assume without loss of generality that $a_2, a_3, a_4$ is negative while $a_1$ is positive. Combining this with the normalization condition $a_1 + a_2 + a_3 + a_4 = 1$ means that $a_1$ must be greater than 1. Therefore, $\langle \sigma^z \rangle = a_1 - a_2 $ is greater than one, which is a contradiction. Thus, this case does not arise.

\item \emph{If only one coefficient  is negative --}  Without loss of generality, let us assume only $a_4$ is negative. In this case, we adopt the strategy of showing that there for every such decomposition, either the state is a stabilizer state or there always exists another decomposition with two negative coefficients which leads to a lower RoM - hence this choice can not be an optimal decomposition. To this end, let us first choose a $\mu \in (0,1) $ such that $a_2  + \mu a_4 = - \epsilon$, where $\epsilon$ is an arbitrarily small positive real number. If no such $\mu$ can be found in this range, then choose a $\mu \geq 1$ for which  $a_2  + \mu a_4 = 0 $ Thus, since $a_1 > a_2$, it is always possible to choose $\epsilon$ in such a way that $a_1 + \mu a_4$ is a positive number. Now, let us note that $|0\rangle \langle 0| + |1\rangle \langle 1|= |+\rangle \langle +| + |-\rangle \langle -| $, substituting this in the assumed optimal decomposition yields the following decomposition for $\rho$, $\rho = (a_1 + \mu a_4) |0\rangle \langle 0| + (a_2 + \mu a_4) |1\rangle \langle 1| + (a_3 - \mu a_4) |+\rangle \langle + | + (1-\mu) a_4 |-\rangle \langle -|.$

Now, let us note that only the coefficients of $|1\rangle\langle 1|$, and $|-\rangle\langle -|$ may be negative here. If $\mu < 1$,  then the new decomposition with two negative coefficients leads to a lower RoM value of $2|1-\mu| + 2 \epsilon$, which is less than the assumed optimal decomposition if $\epsilon$ is arbitrarily small. If $\mu \geq 1$, then the state lies within the stabilizer polytope since all the coefficients are now positive. Hence the proof is complete.

\item \emph{If two of the coefficients are negative --}  Without loss of generality, we assume $a_1$, and $a_3$ are both positive, while $a_2$, and $a_4$ are both negative. Hence, the RoM is given by $ -2 a_2  - 2a_4 = (a_1-  a_2) - (a_1 + a_2) + (a_3 - a_4) - (a_3 + a_4) = \langle \sigma^z \rangle + \langle \sigma^x \rangle -1$.
\end{enumerate}

Thus, the expression in the main text (Eq.~4), that RoM equals $\max[ \langle \sigma^x \rangle + \langle \sigma^z \rangle -1, 0]$, is proved. 

\section{Scaling of transverse magnetisation}
\label{Appendix B}

\begin{figure}[t]
\centering
\includegraphics[width = 0.4\linewidth]{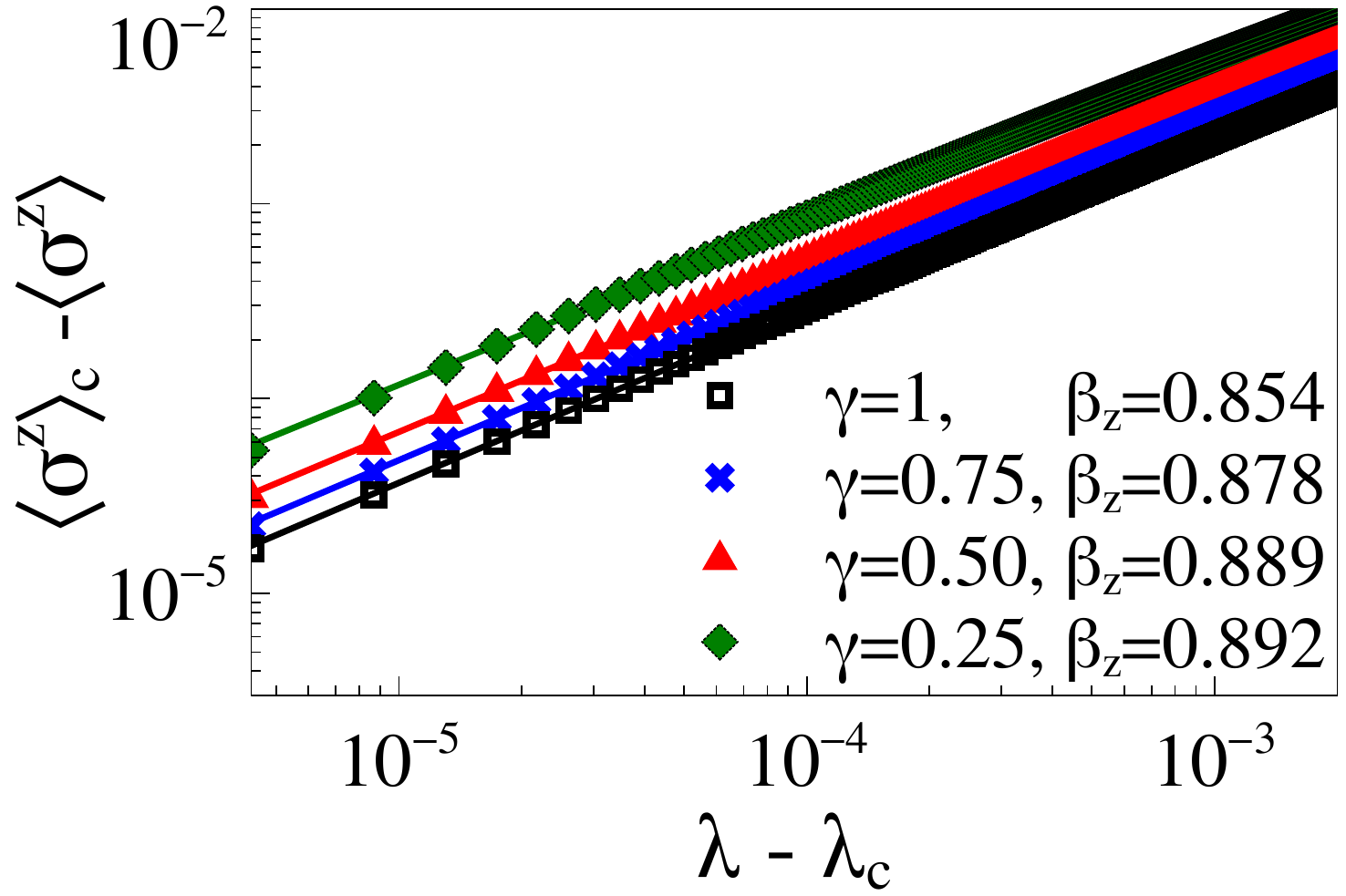}
\caption{Scaling of deviation of transverse magnetization $\langle \sigma^z \rangle$ from its magnitude at $\lambda_c$, with deviation from criticality $\lambda - \lambda_c$ for an infinite transverse XY chain with different $\gamma$ (points), and the exponent $\beta_z$ extracted from the corresponding linear fits (solid straight lines).  }
\label{plot_supp}
\end{figure}

In this subsection, we present numerical evidence that the transverse magnetization has an algebraic behaviour near criticality, i.e., \begin{equation}
\langle \sigma^{z} \rangle \simeq \langle \sigma^{z} \rangle_c + K_z (\lambda - \lambda_c)^{\beta_z} \,.
\end{equation}

As is depicted in figure~\ref{plot_supp}, the algebraic behaviour of transverse magnetization close to the critical point is clear. The scaling exponent $\beta_z$ broadly lies in the range $\in (0.8, 0.9)$. The detailed values of scaling exponents $\beta_z$ for different values of anisotropy, along with the fitting lines are shown in the corresponding figure here.

\section*{References}
\bibliographystyle{iopart-num}  
\bibliography{Spin_Chain_Magic_Draft}

\providecommand{\newblock}{}
\begin{thebibliography}{10}
\expandafter\ifx\csname url\endcsname\relax
  \def\url#1{{\tt #1}}\fi
\expandafter\ifx\csname urlprefix\endcsname\relax\def\urlprefix{URL }\fi
\providecommand{\eprint}[2][]{\url{#2}}

\bibitem{sachdev2007quantum}
Sachdev S 2007 {\em Handbook of Magnetism and Advanced Magnetic Materials\/}

\bibitem{many_body_rmp}
Amico L, Fazio R, Osterloh A and Vedral V 2008 {\em Rev. Mod. Phys.\/}
  \href{http://dx.doi.org/10.1103/RevModPhys.80.517}{{\bf 80}(2) 517--576}

\bibitem{bose2003quantum}
Bose S 2003 {\em Phys. Rev. Lett.\/}
  \href{http://dx.doi.org/10.1103/PhysRevLett.91.207901}{{\bf 91}(20) 207901}

\bibitem{bose2007quantum}
Bose S 2007 {\em Contemporary Physics\/}
  \href{http://dx.doi.org/10.1080/00107510701342313}{{\bf 48} 13--30}

\bibitem{zanardi2008quantum}
Zanardi P, Paris M~G~A and Campos~Venuti L 2008 {\em Phys. Rev. A\/}
  \href{http://dx.doi.org/10.1103/PhysRevA.78.042105}{{\bf 78}(4) 042105}

\bibitem{invernizzi2008optimal}
Invernizzi C, Korbman M, Campos~Venuti L and Paris M~G~A 2008 {\em Phys. Rev.
  A\/} \href{http://dx.doi.org/10.1103/PhysRevA.78.042106}{{\bf 78}(4) 042106}

\bibitem{frerot2018quantum}
Fr\'erot I and Roscilde T 2018 {\em Phys. Rev. Lett.\/}
  \href{http://dx.doi.org/10.1103/PhysRevLett.121.020402}{{\bf 121}(2) 020402}

\bibitem{RevModPhys.89.035002}
Degen C~L, Reinhard F and Cappellaro P 2017 {\em Rev. Mod. Phys.\/}
  \href{http://dx.doi.org/10.1103/RevModPhys.89.035002}{{\bf 89}(3) 035002}

\bibitem{yao2011robust}
Yao N~Y, Jiang L, Gorshkov A~V, Gong Z~X, Zhai A, Duan L~M and Lukin M~D 2011
  {\em Phys. Rev. Lett.\/}
  \href{http://dx.doi.org/10.1103/PhysRevLett.106.040505}{{\bf 106}(4) 040505}

\bibitem{banchi2011nonperturbative}
Banchi L, Bayat A, Verrucchi P and Bose S 2011 {\em Phys. Rev. Lett.\/}
  \href{http://dx.doi.org/10.1103/PhysRevLett.106.140501}{{\bf 106}(14) 140501}

\bibitem{correa2013performance}
Correa L~A, Palao J~P, Adesso G and Alonso D 2013 {\em Phys. Rev. E\/}
  \href{http://dx.doi.org/10.1103/PhysRevE.87.042131}{{\bf 87}(4) 042131}

\bibitem{PhysRevA.97.042124}
Mohammady M~H, Choi H, Trusheim M~E, Bayat A, Englund D and Omar Y 2018 {\em
  Phys. Rev. A\/} \href{http://dx.doi.org/10.1103/PhysRevA.97.042124}{{\bf
  97}(4) 042124}

\bibitem{Osterloh2002}
Osterloh A O~F~G and Fazio R 2002 {\em Nature\/}
  \href{http://dx.doi.org/10.1038/416608a}{{\bf 416}(6881) 608}

\bibitem{Osborne2002}
Osborne T~J and Nielsen M~A 2002 {\em Phys. Rev. A\/}
  \href{http://dx.doi.org/10.1103/PhysRevA.66.032110}{{\bf 66}(3) 032110}

\bibitem{vidal2003entanglement}
Vidal G, Latorre J~I, Rico E and Kitaev A 2003 {\em Phys. Rev. Lett.\/}
  \href{http://dx.doi.org/10.1103/PhysRevLett.90.227902}{{\bf 90}(22) 227902}

\bibitem{bayat2012entanglement}
Bayat A, Bose S, Sodano P and Johannesson H 2012 {\em Phys. Rev. Lett.\/}
  \href{http://dx.doi.org/10.1103/PhysRevLett.109.066403}{{\bf 109}(6) 066403}

\bibitem{bayat2017scaling}
Bayat A 2017 {\em Phys. Rev. Lett.\/}
  \href{http://dx.doi.org/10.1103/PhysRevLett.118.036102}{{\bf 118}(3) 036102}

\bibitem{sun2010fisher}
Sun Z, Ma J, Lu X~M and Wang X 2010 {\em Phys. Rev. A\/}
  \href{http://dx.doi.org/10.1103/PhysRevA.82.022306}{{\bf 82}(2) 022306}

\bibitem{le2007entanglement}
Le~Hur K, Doucet-Beaupr\'e P and Hofstetter W 2007 {\em Phys. Rev. Lett.\/}
  \href{http://dx.doi.org/10.1103/PhysRevLett.99.126801}{{\bf 99}(12) 126801}

\bibitem{sorensen2007quantum}
S{\o}rensen E~S, Chang M~S, Laflorencie N and Affleck I 2007 {\em Journal of
  Statistical Mechanics: Theory and Experiment\/}
  \href{http://dx.doi.org/10.1088/1742-5468/2007/08/p08003}{{\bf 2007}
  P08003--P08003}

\bibitem{bayat2014order}
Bayat A, Johannesson H, Bose S and Sodano P 2014 {\em Nature Communications\/}
  \href{http://dx.doi.org/10.1038/ncomms4784}{{\bf 5} 3784}

\bibitem{AdessoPRL2008}
Giampaolo S~M, Adesso G and Illuminati F 2008 {\em Phys. Rev. Lett.\/}
  \href{http://dx.doi.org/10.1103/PhysRevLett.100.197201}{{\bf 100}(19) 197201}

\bibitem{AdessoPRB2009}
Giampaolo S~M, Adesso G and Illuminati F 2009 {\em Phys. Rev. B\/}
  \href{http://dx.doi.org/10.1103/PhysRevB.79.224434}{{\bf 79}(22) 224434}

\bibitem{AmicoEPL2011}
{Tomasello} B, {Rossini} D, {Hamma} A and {Amico} L 2011 {\em EPL (Europhysics
  Letters)\/} \href{http://dx.doi.org/10.1209/0295-5075/96/27002}{{\bf 96}
  27002}

\bibitem{Vedral2012}
{Cui} J, {Gu} M, {Kwek} L~C, {Santos} M~F, {Fan} H and {Vedral} V 2012 {\em
  Nature Communications\/} \href{http://dx.doi.org/10.1038/ncomms1809}{{\bf 3}
  812}

\bibitem{horodecki2001entanglement}
Horodecki M 2001 {\em Quantum Information \& Computation\/} {\bf 1} 3--26

\bibitem{plenio2014introduction}
Plenio M~B and Virmani S~S 2014 An introduction to entanglement theory {\em
  Quantum Information and Coherence\/} (Springer) pp 173--209

\bibitem{Amico2007}
Amico L and Patan{\`{e}} D 2007 {\em Europhysics Letters ({EPL})\/}
  \href{http://dx.doi.org/10.1209/0295-5075/77/17001}{{\bf 77} 17001}

\bibitem{bravyi}
Bravyi S and Kitaev A 2005 {\em Phys. Rev. A\/}
  \href{http://dx.doi.org/10.1103/PhysRevA.71.022316}{{\bf 71}(2) 022316}

\bibitem{Ollivier2001}
Ollivier H and Zurek W~H 2001 {\em Phys. Rev. Lett.\/}
  \href{http://dx.doi.org/10.1103/PhysRevLett.88.017901}{{\bf 88}(1) 017901}

\bibitem{Henderson2001}
Henderson L and Vedral V 2001 {\em Journal of Physics A: Mathematical and
  General\/} \href{http://dx.doi.org/10.1088/0305-4470/34/35/315}{{\bf 34}
  6899--6905}

\bibitem{baumgratz}
Baumgratz T, Cramer M and Plenio M~B 2014 {\em Phys. Rev. Lett.\/}
  \href{http://dx.doi.org/10.1103/PhysRevLett.113.140401}{{\bf 113}(14) 140401}

\bibitem{Marvian2014}
{Marvian} I and {Spekkens} R~W 2014 {\em Nature Communications\/}
  \href{http://dx.doi.org/10.1038/ncomms4821}{{\bf 5} 3821}

\bibitem{njp_magic}
{Veitch} V, {Hamed Mousavian} S~A, {Gottesman} D and {Emerson} J 2014 {\em New
  Journal of Physics\/}
  \href{http://dx.doi.org/10.1088/1367-2630/16/1/013009}{{\bf 16} 013009}

\bibitem{goursanders}
Ahmadi M, Dang H~B, Gour G and Sanders B~C 2018 {\em Phys. Rev. A\/}
  \href{http://dx.doi.org/10.1103/PhysRevA.97.062332}{{\bf 97}(6) 062332}

\bibitem{campbell}
Campbell E and Browne D 2010 {\em Phys. Rev. Lett.\/}
  \href{http://dx.doi.org/10.1103/PhysRevLett.104.030503}{{\bf 104}(3) 030503}

\bibitem{howard}
Howard M and Campbell E 2017 {\em Phys. Rev. Lett.\/}
  \href{http://dx.doi.org/10.1103/PhysRevLett.118.090501}{{\bf 118}(9) 090501}

\bibitem{entmagic}
T\'oth G and G\"uhne O 2005 {\em Phys. Rev. A\/}
  \href{http://dx.doi.org/10.1103/PhysRevA.72.022340}{{\bf 72}(2) 022340}

\bibitem{entmagic2}
Audenaert K~M~R and Plenio M~B 2005 {\em New Journal of Physics\/}
  \href{http://dx.doi.org/10.1088/1367-2630/7/1/170}{{\bf 7} 170--170}

\bibitem{coherence_magical}
Mukhopadhyay C, Sazim S and Pati A~K 2018 {\em Journal of Physics A:
  Mathematical and Theoretical\/}
  \href{http://dx.doi.org/10.1088/1751-8121/aac8e8}{{\bf 51} 414006}

\bibitem{cont_vs_magic}
Zhan X, Zhang X, Li J, Zhang Y, Sanders B~C and Xue P 2016 {\em Phys. Rev.
  Lett.\/} \href{http://dx.doi.org/10.1103/PhysRevLett.116.090401}{{\bf 116}(9)
  090401}

\bibitem{contextuality_supplies_magic}
{Howard} M, {Wallman} J, {Veitch} V and {Emerson} J 2014 {\em Nature\/}
  \href{http://dx.doi.org/10.1038/nature13460}{{\bf 510} 351--355}

\bibitem{nong1}
Albarelli F, Genoni M~G, Paris M~G~A and Ferraro A 2018 {\em Phys. Rev. A\/}
  \href{http://dx.doi.org/10.1103/PhysRevA.98.052350}{{\bf 98}(5) 052350}

\bibitem{nong2}
Takagi R and Zhuang Q 2018 {\em Phys. Rev. A\/}
  \href{http://dx.doi.org/10.1103/PhysRevA.97.062337}{{\bf 97}(6) 062337}

\bibitem{ferrie}
Veitch V, Ferrie C, Gross D and Emerson J 2012 {\em New Journal of Physics\/}
  \href{http://dx.doi.org/10.1088/1367-2630/14/11/113011}{{\bf 14} 113011}

\bibitem{marie}
Mari A and Eisert J 2012 {\em Phys. Rev. Lett.\/}
  \href{http://dx.doi.org/10.1103/PhysRevLett.109.230503}{{\bf 109}(23) 230503}

\bibitem{jones}
Jones C 2013 {\em Phys. Rev. A\/}
  \href{http://dx.doi.org/10.1103/PhysRevA.87.042305}{{\bf 87}(4) 042305}

\bibitem{du}
Zheng W, Yu Y, Pan J, Zhang J, Li J, Li Z, Suter D, Zhou X, Peng X and Du J
  2015 {\em Phys. Rev. A\/}
  \href{http://dx.doi.org/10.1103/PhysRevA.91.022314}{{\bf 91}(2) 022314}

\bibitem{reedmueller}
Campbell E~T, Anwar H and Browne D~E 2012 {\em Phys. Rev. X\/}
  \href{http://dx.doi.org/10.1103/PhysRevX.2.041021}{{\bf 2}(4) 041021}

\bibitem{factory}
O'Gorman J and Campbell E~T 2017 {\em Phys. Rev. A\/}
  \href{http://dx.doi.org/10.1103/PhysRevA.95.032338}{{\bf 95}(3) 032338}

\bibitem{unified}
Campbell E~T and Howard M 2017 {\em Phys. Rev. A\/}
  \href{http://dx.doi.org/10.1103/PhysRevA.95.022316}{{\bf 95}(2) 022316}

\bibitem{bravyi_haah}
Bravyi S and Haah J 2012 {\em Phys. Rev. A\/}
  \href{http://dx.doi.org/10.1103/PhysRevA.86.052329}{{\bf 86}(5) 052329}

\bibitem{hastings_haah}
Hastings M~B and Haah J 2018 {\em Phys. Rev. Lett.\/}
  \href{http://dx.doi.org/10.1103/PhysRevLett.120.050504}{{\bf 120}(5) 050504}

\bibitem{krishna_tillich}
Krishna A and Tillich J~P 2019 {\em Phys. Rev. Lett.\/}
  \href{http://dx.doi.org/10.1103/PhysRevLett.123.070507}{{\bf 123}(7) 070507}

\bibitem{magcat}
Campbell E~T 2011 {\em Phys. Rev. A\/}
  \href{http://dx.doi.org/10.1103/PhysRevA.83.032317}{{\bf 83}(3) 032317}

\bibitem{howard_tight}
Dawkins H and Howard M 2015 {\em Phys. Rev. Lett.\/}
  \href{http://dx.doi.org/10.1103/PhysRevLett.115.030501}{{\bf 115}(3) 030501}

\bibitem{Lieb1961}
Lieb E, Schultz T and Mattis D 1961 {\em Annals of Physics\/}
  \href{http://dx.doi.org/https://doi.org/10.1016/0003-4916(61)90115-4}{{\bf
  16} 407 -- 466} ISSN 0003-4916

\bibitem{Dutta2010}
Dutta A, Aeppli G, Chakrabarti B~K, Divakaran U, Rosenbaum T~F and Sen D 2015
  {\em Quantum Phase Transitions in Transverse Field Spin Models: From
  Statistical Physics to Quantum Information\/} (Cambridge University Press)

\bibitem{Barouch1970}
Barouch E, McCoy B~M and Dresden M 1970 {\em Phys. Rev. A\/}
  \href{http://dx.doi.org/10.1103/PhysRevA.2.1075}{{\bf 2}(3) 1075--1092}

\bibitem{Barouch1971}
Barouch E and McCoy B~M 1971 {\em Phys. Rev. A\/}
  \href{http://dx.doi.org/10.1103/PhysRevA.3.786}{{\bf 3}(2) 786--804}

\bibitem{Pfeuty1970}
Pfeuty P 1970 {\em Annals of Physics\/}
  \href{http://dx.doi.org/https://doi.org/10.1016/0003-4916(70)90270-8}{{\bf
  57} 79 -- 90} ISSN 0003-4916

\bibitem{lidar_book}
Lidar D and Brun T 2013 {\em Quantum Error Correction\/} (Cambridge University
  Press)

\bibitem{cluster}
Briegel H~J 2009 {\em Cluster States\/} (Berlin, Heidelberg: Springer Berlin
  Heidelberg) pp 96--105 ISBN 978-3-540-70626-7

\bibitem{gk}
Gottesman D 1998 {\em arXiv:quant-ph/9807006\/}

\bibitem{heinrich}
Heinrich M and Gross D 2019 {\em {Quantum}\/}
  \href{http://dx.doi.org/10.22331/q-2019-04-08-132}{{\bf 3} 132} ISSN
  2521-327X

\bibitem{White1992}
White S~R 1992 {\em Phys. Rev. Lett.\/}
  \href{http://dx.doi.org/10.1103/PhysRevLett.69.2863}{{\bf 69}(19) 2863--2866}

\bibitem{Pirvu2010}
Pirvu B, Murg V, Cirac J~I and Verstraete F 2010 {\em New Journal of Physics\/}
  \href{http://dx.doi.org/10.1088/1367-2630/12/2/025012}{{\bf 12} 025012}

\bibitem{Schollwock2011}
Schollw\"{o}ck U 2011 {\em Ann. Phys.\/}
  \href{http://dx.doi.org/10.1016/j.aop.2010.09.012}{{\bf 326} 96--192}

\bibitem{ITensor}
 {\em \mbox{ITensor Library} (version 3.0.0) http://itensor.org\/}

\bibitem{Barber1983}
Barber M 1983 {\em Finite size scaling\/} ({\em Phase Transitions and critical
  phenomena, edited by C. Domb and J.L. Leibovitz\/} vol~8) (Academic Press,
  London)

\bibitem{McCulloch2008}
{McCulloch} I~P 2008 {\em arXiv e-prints\/} arXiv:0804.2509

\bibitem{cvx1}
Grant M and Boyd S 2014 {CVX}: Matlab software for disciplined convex
  programming, version 2.1

\bibitem{cvx2}
Grant M and Boyd S 2008 Graph implementations for nonsmooth convex programs
  {\em Recent Advances in Learning and Control\/} Lecture Notes in Control and
  Information Sciences ed Blondel V, Boyd S and Kimura H (Springer-Verlag
  Limited) pp 95--110

\bibitem{Radhakrishnan2016}
Radhakrishnan C, Parthasarathy M, Jambulingam S and Byrnes T 2016 {\em Phys.
  Rev. Lett.\/} \href{http://dx.doi.org/10.1103/PhysRevLett.116.150504}{{\bf
  116}(15) 150504}

\bibitem{Dillenschneider2008}
Dillenschneider R 2008 {\em Phys. Rev. B\/}
  \href{http://dx.doi.org/10.1103/PhysRevB.78.224413}{{\bf 78}(22) 224413}

\bibitem{Sarandy2009}
Sarandy M~S 2009 {\em Phys. Rev. A\/}
  \href{http://dx.doi.org/10.1103/PhysRevA.80.022108}{{\bf 80}(2) 022108}

\bibitem{Krutitsky2017}
Krutitsky K~V, Osterloh A and Sch{\"u}tzhold R 2017 {\em Scientific Reports\/}
  \href{http://dx.doi.org/10.1038/s41598-017-03402-8}{{\bf 7} 3634} ISSN
  2045-2322

\bibitem{Sachdev1999}
Sachdev S 1999 {\em Quantum phase transitions\/} (Cambridge University Press,
  Cambridge)

\bibitem{Frerot2019}
{Fr{\'e}rot} I and {Roscilde} T 2019 {\em Nature Communications\/}
  \href{http://dx.doi.org/10.1038/s41467-019-08324-9}{{\bf 10} 577}

\bibitem{Continentino2001}
Continentino M 2001 {\em Quantum Scaling in Many Body Systems\/} (World
  Scientific, Singapore)

\bibitem{Zhu2003}
Zhu L, Garst M, Rosch A and Si Q 2003 {\em Phys. Rev. Lett.\/}
  \href{http://dx.doi.org/10.1103/PhysRevLett.91.066404}{{\bf 91}(6) 066404}

\bibitem{Maziero2010}
Maziero J, Guzman H~C, C\'eleri L~C, Sarandy M~S and Serra R~M 2010 {\em Phys.
  Rev. A\/} \href{http://dx.doi.org/10.1103/PhysRevA.82.012106}{{\bf 82}(1)
  012106}

\bibitem{Maziero2012}
Maziero J, Céleri L, Serra R and Sarandy M 2012 {\em Physics Letters A\/}
  \href{http://dx.doi.org/https://doi.org/10.1016/j.physleta.2012.03.029}{{\bf
  376} 1540 -- 1544} ISSN 0375-9601

\bibitem{Auccaise2011}
Auccaise R, C\'eleri L~C, Soares-Pinto D~O, deAzevedo E~R, Maziero J, Souza
  A~M, Bonagamba T~J, Sarthour R~S, Oliveira I~S and Serra R~M 2011 {\em Phys.
  Rev. Lett.\/} \href{http://dx.doi.org/10.1103/PhysRevLett.107.140403}{{\bf
  107}(14) 140403}

\bibitem{Cornelio2012}
Cornelio M~F, Far\'{\i}as O~J, Fanchini F~F, Frerot I, Aguilar G~H, Hor-Meyll
  M~O, de~Oliveira M~C, Walborn S~P, Caldeira A~O and Ribeiro P~H~S 2012 {\em
  Phys. Rev. Lett.\/}
  \href{http://dx.doi.org/10.1103/PhysRevLett.109.190402}{{\bf 109}(19) 190402}

\bibitem{Paula2013}
Paula F~M, Silva I~A, Montealegre J~D, Souza A~M, deAzevedo E~R, Sarthour R~S,
  Saguia A, Oliveira I~S, Soares-Pinto D~O, Adesso G and Sarandy M~S 2013 {\em
  Phys. Rev. Lett.\/}
  \href{http://dx.doi.org/10.1103/PhysRevLett.111.250401}{{\bf 111}(25) 250401}

\bibitem{Wang2019}
Wang X, Wilde M~M and Su Y 2019 {\em New Journal of Physics\/}
  \href{http://dx.doi.org/10.1088/1367-2630/ab451d}{{\bf 21} 103002}

\end{thebibliography}

\end{document}